\documentclass[a4paper,11pt]{article}

\pdfoutput=1 

\usepackage{jheppub} 
\usepackage[dvipsnames]{xcolor}
\usepackage{tikz}
\usetikzlibrary{patterns,arrows,decorations.pathmorphing,decorations.markings}

\usepackage{slashed,adjustbox,bm,mathrsfs,bbm}
\usepackage{subcaption}
\usepackage{cleveref}
\usepackage[T1]{fontenc} 
\usepackage[utf8]{inputenc}
\usepackage{mathtools}

\newcommand{\cA}{\mathcal{A}}
\newcommand{\cM}{\mathcal{M}}
\newcommand{\cT}{\mathcal{T}}

\newcommand{\cD}{\mathcal{D}}
\newcommand{\cL}{\mathcal{L}}

\newcommand{\cY}{\mathcal{Y}}

\def\rd{{\rm d}}

\def\coef#1#2#3{ {\vphantom{A^2}}^{#1}\!{{C}}^{#3}_{#2} }
\def\dcoef#1#2#3{ {\vphantom{A^2}}^{#1}\!{\dot{{C}}}^{#3}_{#2} }

\def\op#1#2#3{ {\vphantom{A^2}}^{#1}\!{Q}^{#3}_{#2} }

\def\an#1{\langle #1 \rangle}

\newcommand{\kappaHypKinSix}{\kappa_{1}}
\newcommand{\kappaPauliKinSix}{\kappa_{2}}
\newcommand{\kappaHypKinEight}{\kappa_{1}^{(8)}}
\newcommand{\kappaPauliKinEight}{\kappa_{2}^{(8)}}
\newcommand{\kappaPauliKinEightTwo}{\kappa_{3}}
\newcommand{\kappaKinSixKinSix}{\kappa_{4}}
\newcommand{\kappaKinSixKinSixSUTwo}{\kappa_{5}}
\newcommand{\kappaKinSixKinSixud}{\kappa_{6}}
\newcommand{\kappaYukawaFourYukawaSix}{\kappa_{7}}
\newcommand{\kappaYukawaFourYukawaEight}{\kappa_{7}^{(8)}}
\newcommand{\kappaYukawaSixYukawaSix}{\kappa_{8}}
\newcommand{\kappaYukawaKinSix}{\kappa_{9}}
\newcommand{\kappaYukawaKinEight}{\kappa_{9}^{(8)}}
\newcommand{\kappaYukawaKinThreeSix}{\kappa_{10}}
\newcommand{\kappaYukawaKinThreeEight}{\kappa_{10}^{(8)}}
\newcommand{\kappaYukawaKinudSix}{\kappa_{11}}
\newcommand{\kappaYukawaKinudEight}{\kappa_{11}^{(8)}}
\newcommand{\kappaYukawaKinTwoEight}{\kappa_{12}}
\newcommand{\kappaYukawaKinSixKinSix}{\kappa_{13}}
\newcommand{\kappaYukawaKinSixKinSixThree}{\kappa_{14}}
\newcommand{\kappaYukawaKinSixThreeKinSixThree}{\kappa_{15}}
\newcommand{\kappaYukawaSixKinSix}{\kappa_{16}}
\newcommand{\kappaYukawaSixKinSixThree}{\kappa_{17}}
\newcommand{\kappaYukawaSixKinSixud}{\kappa_{18}}
\newcommand{\kappaYukawaKinSixudKinSixud}{\kappa_{19}}
\newcommand{\kappaYukawaKinSixThreeKinSixud}{\kappa_{20}}
\newcommand{\kappaYukawaYYYyukawaSix}{\kappa_{21}}
\newcommand{\kappaYukawaYYYyukawaEight}{\kappa_{21}^{(8)}}
\newcommand{\kappaYukawaYYyukawaSixYukawaSix}{\kappa_{22}}

\newcommand{\tauGG}{\tau_{0}}
\newcommand{\tauBB}{\tau_{1}}
\newcommand{\tauBW}{\tau_{2}}
\newcommand{\tauBWp}{\tau_{2}^{\prime}}
\newcommand{\tauBWpp}{\tau_{2}^{\prime\prime}}
\newcommand{\tauWW}{\tau_{3}}
\newcommand{\tauWWp}{\tau_{3}^{\prime}}
\newcommand{\tauYukDipoleBBSix}{\tau_{4}}
\newcommand{\tauYukDipoleBBEight}{\tau_{4}^{(8)}}
\newcommand{\tauYukDipoleBBtildeSix}{\tau_{5}}
\newcommand{\tauYukDipoleBBtildeEight}{\tau_{5}^{(8)}}
\newcommand{\tauYukDipoleBWSix}{\tau_{6}}
\newcommand{\tauYukDipoleBWEight}{\tau_{6}^{(8)}}
\newcommand{\tauYukDipoleBWtildeSix}{\tau_{7}}
\newcommand{\tauYukDipoleBWtildeEight}{\tau_{7}^{(8)}}
\newcommand{\tauYukDipoleWBSix}{\tau_{8}}
\newcommand{\tauYukDipoleWBEight}{\tau_{8}^{(8)}}
\newcommand{\tauYukDipoleWtildeBSix}{\tau_{9}}
\newcommand{\tauYukDipoleWtildeBEight}{\tau_{9}^{(8)}}
\newcommand{\tauYukDipoleWWSix}{\tau_{10}}
\newcommand{\tauYukDipoleWWEight}{\tau_{10}^{(8)}}
\newcommand{\tauYukDipoleWtildeWSix}{\tau_{11}}
\newcommand{\tauYukDipoleWtildeWEight}{\tau_{11}^{(8)}}
\newcommand{\tauYukDipoleWWtwoEight}{\tau_{12}}
\newcommand{\tauYukDipoleWtildeWtwoEight}{\tau_{13}}
\newcommand{\tauYukDipoleGGSix}{\tau_{14}}
\newcommand{\tauYukDipoleGGEight}{\tau_{14}^{(8)}}
\newcommand{\tauYukDipoleGGtildeSix}{\tau_{15}}
\newcommand{\tauYukDipoleGGtildeEight}{\tau_{15}^{(8)}}
\newcommand{\tauYukKinSixDipoleB}{\tau_{16}}
\newcommand{\tauYukKinSixDipoleBtilde}{\tau_{17}}
\newcommand{\tauYukKinSixDipoleW}{\tau_{18}}
\newcommand{\tauYukKinSixDipoleWtilde}{\tau_{19}}
\newcommand{\tauYukKinThreeSixDipoleB}{\tau_{20}}
\newcommand{\tauYukKinThreeSixDipoleBtilde}{\tau_{21}}
\newcommand{\tauYukKinThreeSixDipoleW}{\tau_{22}}
\newcommand{\tauYukKinThreeSixDipoleWtilde}{\tau_{23}}
\newcommand{\tauYukKinThreeSixDipoleWeps}{\tau_{24}}
\newcommand{\tauYukKinThreeSixDipoleWtildeeps}{\tau_{25}}
\newcommand{\tauYukKinudSixDipoleW}{\tau_{26}}
\newcommand{\tauYukKinudSixDipoleWtilde}{\tau_{27}}
\newcommand{\tauYukKinudSixDipoleWeps}{\tau_{28}}
\newcommand{\tauYukKinudSixDipoleWtildeeps}{\tau_{29}}

\newcommand{\tauYukTwoDipoleBB}{\tau_{30}}
\newcommand{\tauYukTwoDipoleBBtilde}{\tau_{31}}
\newcommand{\tauYukTwoDipoleBW}{\tau_{32}}
\newcommand{\tauYukTwoDipoleBWtilde}{\tau_{33}}
\newcommand{\tauYukTwoDipoleWW}{\tau_{34}}
\newcommand{\tauYukTwoDipoleWWtilde}{\tau_{35}}
\newcommand{\tauYukTwoDipoleWWtwo}{\tau_{36}}
\newcommand{\tauYukTwoDipoleWWtildetwo}{\tau_{37}}
\newcommand{\tauYukTwoDipoleGG}{\tau_{38}}
\newcommand{\tauYukTwoDipoleGGtilde}{\tau_{39}}

\makeatletter
\newsavebox{\@brx}
\newcommand{\llangle}[1][]{\savebox{\@brx}{\(\m@th{#1\langle}\)}%
  \mathopen{\copy\@brx\kern-0.5\wd\@brx\usebox{\@brx}}}
\newcommand{\rrangle}[1][]{\savebox{\@brx}{\(\m@th{#1\rangle}\)}%
  \mathclose{\copy\@brx\kern-0.5\wd\@brx\usebox{\@brx}}}
\makeatother

\subheader{\normalsize
	\vspace*{-3.7em}
	\begin{flushright}
		CALT-TH-2023-024\\
            FERMILAB-PUB-23-362-T
	\end{flushright}
}

\title{\boldmath Fermion Geometry and the Renormalization of the Standard Model Effective Field Theory}

\author[a]{Benoît Assi,}

\author[b]{Andreas Helset,}

\author[c]{Aneesh V.~Manohar,}

\author[c]{Julie Pag\`es,}

\author[c]{Chia-Hsien Shen}

\affiliation[a]{Fermi National Accelerator Laboratory, Batavia, IL, 60510,  USA}

\affiliation[b]{Walter Burke Institute for Theoretical Physics,
California Institute of Technology,\\ Pasadena, CA 91125, USA}

\affiliation[c]{Physics Department,
University of California San Diego, 9500 Gilman Drive,\\ La Jolla, CA 92093-0319, USA}

\emailAdd{bassi@fnal.gov}
\emailAdd{ahelset@caltech.edu}
\emailAdd{amanohar@ucsd.edu}
\emailAdd{jcpages@ucsd.edu}
\emailAdd{c9shen@ucsd.edu}

\abstract{The geometry of field space governs on-shell scattering amplitudes. We formulate a geometric description of effective field theories which extends previous results for scalars and gauge fields to fermions. The field-space geometry reorganizes and simplifies the computation of quantum  loop corrections.  Using this geometric framework, we calculate the fermion loop contributions to the renormalization group equations for bosonic operators in the Standard Model Effective Field Theory up to mass dimension eight.}

\allowdisplaybreaks

\begin{document} 
\maketitle
\flushbottom

\section{Introduction}\label{sec:Introduction}

Effective field theory (EFT) is a quantum field theory with a systematic expansion in a hierarchy of scales. 
There is an inherent redundancy in the Lagrangian description of effective field theories: fields can be redefined while leaving physical observables unchanged \cite{Chisholm:1961tha,Kamefuchi:1961sb,Politzer:1980me,Arzt:1993gz}. The freedom of applying field redefinitions sometimes obscures the underlying simplicity in physical EFT observables. 
Field redefinitions can be viewed as coordinate changes on a field-space manifold. Since $S$-matrix elements are invariant under field redefinitions, they are invariant under coordinate changes of the field-space manifold, and only depend on geometric quantities. The connection between field redefinitions and geometry is well-established for scalar field redefinitions of the form $\phi^{I} \rightarrow \phi^{\prime I}(\phi)$ \cite{Meetz:1969as,Honerkamp:1971xtx,Honerkamp:1971sh}. This geometric picture has several practical consequences, e.g., that scattering amplitudes for scalars are given in terms of the Riemann curvature and covariant derivatives \cite{Volkov:1970aa}, and that the scattering amplitudes satisfy a geometric soft theorem \cite{Cheung:2021yog}. The geometric underpinning of EFTs has seen recent interest \cite{Alonso:2015fsp,Alonso:2016oah,Alonso:2017tdy,Helset:2018fgq,Helset:2020yio,Hays:2020scx,Cohen:2020xca,Corbett:2021eux,Corbett:2021cil,Cohen:2021ucp,Martin:2023fad,Gattus:2023gep}, and the geometric framework has been generalized in several directions to include particles with spin and field redefinitions with derivatives \cite{Finn:2020nvn,Cheung:2022vnd,Cohen:2022uuw,Helset:2022tlf,Craig:2023wni}.

The field-space geometry simplifies calculations. For instance, the renormalization group equations (RGEs) for a theory of scalars \cite{Alonso:2016oah} or a theory with scalars and gauge bosons \cite{Helset:2022pde} depend on the curvature in field space. This reorganization of terms into compact geometric objects is particularly useful when the number of effective operators is large.  This is the case for the Standard Model Effective Field Theory (SMEFT), which augments the Standard Model with higher-dimensional operators. The SMEFT RGEs have been calculated for operators with mass dimension six \cite{Jenkins:2013zja,Jenkins:2013wua,Alonso:2013hga}, and partial results are available for operators with mass dimension eight \cite{Chala:2021pll,DasBakshi:2022mwk,AccettulliHuber:2021uoa,Helset:2022pde,DasBakshi:2023htx}. The field-space geometry has been employed to calculate parts of the bosonic RGEs up to mass dimension eight \cite{Helset:2022pde}. In this work, we extend this approach by including the fermion geometry in the calculation of the SMEFT RGEs, and use this to compute one-loop fermionic corrections to bosonic operators  to mass dimension eight.

The paper is organized as follows. In sec.~\ref{sec:Geometry} we define the geometry for the combined scalar-fermion field space. In sec.~\ref{sec:Amplitudes}, we show that scattering amplitudes depend only on geometric quantities, i.e., the curvature in field space and covariant derivatives. In sec.~\ref{sec:Renormalization} we use this field-space geometry to derive a general formula for the fermionic one-loop contribution to the bosonic RGEs. We then apply this formula to the SMEFT in sec.~\ref{sec:smeft} and the Low-energy Effective Field Theory (LEFT) below the electroweak scale in sec.~\ref{sec:LEFT}. We conclude in sec.~\ref{sec:Conclusion}. Appendix~\ref{sec:Operators} lists the operator basis for the SMEFT and the RGE results for the SMEFT are listed in app.~\ref{sec:RGEresults}. Useful identities for selfdual dipoles are summarized in app.~\ref{sec:self-dual}, while the extension of the RGE formula to include Majorana mass and dipole terms is described in app.~\ref{app:Majorana}, which are needed for computing the RGEs in the LEFT~\cite{Jenkins:2017jig,Jenkins:2017dyc}.

\section{Field-space geometry}\label{sec:Geometry}

We start with a theory of scalars, gauge bosons, and fermions, including interactions with at most two derivatives and two fermion fields, ignoring $CP$-violating operators in the bosonic sector for simplicity. The generalization to Lagrangians with higher derivatives is discussed in refs.~\cite{Cheung:2022vnd,Cohen:2022uuw,Craig:2023wni}.
The general Lagrangian takes the form (we use the notation of refs.~\cite{Alonso:2016oah,Helset:2022tlf})
\begin{align}\label{eq:Lagr}\
    \cL =& \frac{1}{2} h_{IJ}(\phi) (D_{\mu} \phi)^{I} (D^{\mu} \phi)^{J} - V(\phi) - \frac{1}{4} g_{AB}(\phi) F^{A}_{\mu\nu} F^{B\mu\nu} 
    \\
    +& \frac{1}{2} i k_{\bar p r}(\phi) \left(\bar \psi^{\bar p} \gamma^{\mu} \overset{\leftrightarrow}{D}_{\mu} \psi^{r}\right) + i \omega_{\bar pr I}(\phi) (D_{\mu} \phi)^{I} \bar \psi^{\bar p} \gamma^{\mu} \psi^{r} - \bar \psi^{\bar p} \cM_{\bar p r}(\phi) \psi^{r} + \bar \psi^{\bar p} \sigma_{\mu\nu} \cT_{\bar p r}^{\mu\nu}(\phi,F) \psi^{r} \,, \nonumber 
\end{align}
where $\bar \psi^{\bar p} \gamma^{\mu} \overset{\leftrightarrow}{D}_{\mu} \psi^{r} = \bar \psi^{\bar p} \gamma^{\mu} \left({D}_{\mu} \psi^{r}\right) - \left(D_{\mu} \bar \psi^{\bar p}\right) \gamma^{\mu}  \psi^{r}$. Here $I,J,K,\dots$ are scalar indices, $A,B,C,\dots$ are gauge field indices, and $p,\bar p,r,\bar r,\dots$ are fermion indices. All quantities $h_{IJ}(\phi)$, $V(\phi)$, $g_{AB}(\phi)$, $k_{\bar p r}(\phi)$, $\omega_{\bar pr I}(\phi)$, and $\cM_{\bar pr}(\phi)$ are functions of the scalar fields, while $\cT^{\mu\nu}_{\bar pr}(\phi,F)$ depends on both the scalar fields and the field strength.
The field strength and covariant derivatives are
\begin{subequations}
\begin{align}
    F^{B}_{\mu\nu} &= \partial_{\mu} A^{B}_{\nu} - \partial_{\nu} A^{B}_{\mu} - f^{B}_{\;\;CD} A^{C}_{\mu} A^{D}_{\nu} \,, \\ 
    (D_{\mu} \phi)^{I} &= \partial_{\mu} \phi^{I} + A^{B}_{\mu} t^{I}_{B}(\phi) \,, \\
    (D_{\mu} \psi)^{p} &= \partial_{\mu} \psi^{p} + A^{B}_{\mu} t^{p}_{B,s} \psi^{s} \,.
\end{align}
\end{subequations}
Note that the gauge couplings have been absorbed into the definitions of $f^{A}_{\;\;BC}$, and the gauge couplings and $i$ into $t^{I}_{B}$ and $t^{p}_{B,s}$ \cite{Helset:2022tlf}. As discussed in ref.~\cite{Alonso:2016oah}, the gauge symmetry is a geometric symmetry of the scalar manifold, and $t^I_B(\phi)$ are Killing vectors, 
\begin{align}
\nabla_J t_{IB}(\phi) + \nabla_I t_{JB}(\phi) = 0 \,,
\end{align}
where the covariant derivative uses the metric connection of $h_{IJ}$ defined below in \cref{eq:ChristoffelScalar}, and $t_{IB}(\phi)=h_{IJ}t^J_{B}(\phi)$ is the Killing vector with the index lowered using the metric. The Killing vectors satisfy the Lie algebra
\begin{align}
\left[t_A,t_B\right] &= f^{C}_{\;\;AB}\, t_C,
\end{align}
where the left-hand side is the Lie bracket of the vectors $t^I_A$. The fermionic generators $t^p_{B,s}$ satisfy the same algebra up to a sign, where the left-hand side is now matrix multiplication.

\subsection{Field redefinitions}\label{sec:field-redef}

Now consider a redefinition of the scalar field, $\phi^{I} \rightarrow \phi^{\prime I}(\phi)$. We will consider only point transformations, i.e., field redefinitions not involving derivatives.
The scalar field transforms as a coordinate under this field redefinition, while $(D_{\mu} \phi)^{I}$ transforms as a vector,
\begin{align}
    (D_{\mu} \phi)^{I} \rightarrow \left(\frac{\partial \phi^{\prime I}}{\partial \phi^{J}} \right) (D_{\mu} \phi)^{J} \,, 
\end{align}
and $h_{IJ}(\phi)$ transforms as a tensor,
\begin{align}
    h_{IJ} \rightarrow \left(\frac{\partial \phi^{K}}{\partial \phi^{\prime I}} \right)\left(\frac{\partial \phi^{L}}{\partial \phi^{\prime J}} \right) h_{KL} \,. 
\end{align}
From this we identify $h_{IJ}(\phi)$ as the metric in the scalar field space, since it has the transformation properties of a metric under coordinate changes.

We want to extend the notion of a field-space geometry to include the fermions. The field-space geometry for gauge fields was discussed in refs.~\cite{Cheung:2022vnd,Helset:2022pde,Helset:2022tlf}, so we drop them in the following discussion for simplicity. In the end, the full field-space geometry will include all scalars, fermions, and gauge bosons. 

The scalar kinetic term involves the scalar field-space metric $h_{IJ}(\phi)$. The gauge kinetic term prefactor $g_{AB}(\phi)$ enters in the metric for the gauge-scalar field space~\cite{Cheung:2022vnd,Helset:2022pde,Helset:2022tlf}.
Thus, it is natural to expect that the prefactor of the fermion kinetic term, $k_{\bar p r}(\phi)$, will play a central role in the field-space geometry for fermions. Under a  redefinition of the fermion field that depends on the scalar field,
\begin{align}
    \psi^{p} \rightarrow R^{p}_{\; s}(\phi) \psi^{s} \,,
    \label{2.7}
\end{align}
the fermion kinetic term prefactor $k_{\bar pr}(\phi)$ mixes with $\omega_{\bar pr I}(\phi)$ as
\begin{align}\label{eq:fieldRedef}
    k_{\bar pr} &\rightarrow \left[ (R^\dagger)^{-1} k R^{-1} \right]_{\bar pr} \,, \\
    \omega_{\bar pr I} &\rightarrow \left[(R^\dagger)^{-1} \omega_{I} R^{-1}\right]_{\bar pr} + \frac{1}{2} \left[ (R^\dagger)^{-1} k (\partial_{I}R^{-1}) \right]_{\bar pr} - \frac{1}{2} \left[(\partial_{I}(R^\dagger)^{-1}) k R^{-1} \right]_{\bar pr} \,, 
\end{align}
where $\partial_I \equiv \partial / \partial \phi^I$. If the fermionic indices $\bar p$ and $r$ are regarded as indices in a local Cartan frame, and eq.~\eqref{2.7} as a frame redefinition, $ k_{\bar pr}$ transforms like the metric and $\omega_{\bar pr I}$ like the Cartan connection. This means that both $k_{\bar pr}$ and $\omega_{\bar pr I}$ are needed in the fermion geometry.

\subsection{Curvature}\label{sec:Curvature}

Before discussing the geometric construction for the fermions, we start with the scalar sector, defined by the metric $h_{IJ}$. From this field-space metric, we can calculate the Christoffel symbol, field-space covariant derivatives, and the Riemann curvature. To distinguish the different geometric objects, all quantities in the combined scalar-fermion field space are denoted with a bar, while no bar is used for the scalar sector. 

The Christoffel symbol is given by
\begin{align}
    \label{eq:ChristoffelScalar}
    \Gamma^{I}_{JK} = \frac{1}{2} h^{IL} \left( h_{LK,J} + h_{JL,K} - h_{JK,L} \right) \,,
\end{align}
where $h_{IJ,K} = \partial_{K} h_{IJ}$. This Christoffel symbol is the connection used in the field-space covariant derivative $\nabla_{I}$. The Riemann curvature is
\begin{align}
    R^{I}_{\;JKL} = \partial_{K} \Gamma^{I}_{LJ} - \partial_{L} \Gamma^{I}_{KJ} + \Gamma^{I}_{KN} \Gamma^{N}_{LJ} - \Gamma^{I}_{LN} \Gamma^{N}_{KJ} \,. 
\end{align}
The curvature is a function of the scalar field $\phi$. It is often useful to evaluate the curvature at the vacuum expectation value (VEV), $\an{\phi^I}=v^{I}$. In particular, the curvature and field-space covariant derivatives appearing in scattering amplitudes and soft theorems are all evaluated at the VEV \cite{Cheung:2021yog}. 

We now combine the space of scalar and fermion fields, and define geometric quantities in this combined space.
Our construction for the scalar-fermion field space is similar to those found in supersymmetric nonlinear sigma models~\cite{Alvarez-Gaume:1981exv} and in refs.~\cite{Finn:2020nvn,Nagai:2021gmo}. The usefulness of the following definitions will be evident in that scattering amplitudes and renormalization group equations depend on geometric quantities in this combined space.

The first difference from the scalar case above is that we must include the fermions as coordinates on the manifold. The natural extension of a Riemannian manifold which allows for Grassmann coordinates is called a {\it supermanifold} \cite{DeWitt:2012mdz,rogers2007supermanifolds}. This is not limited to theories with supersymmetry, and our main application in this work is the SMEFT (which is not supersymmetric). All geometric quantities in the scalar-fermion field space inherit properties from the underlying supermanifold. In essence, this means that the ordering of terms and operations like differentiation need to be handled with care, since some of the coordinates are anticommuting. 

Similar to the scalar case, we will evaluate expressions at the VEV. When we evaluate expressions at the VEV, we assume that the VEVs for the fermions vanish. Note that even though the fermions have vanishing VEV, derivatives with respect to the fermion field can still be nonzero, so the VEV is only taken after evaluating all derivatives.

We start with the metric for the scalar-fermion field space evaluated at the VEV,
\begin{align}
    \bar g_{ab}\left(\langle \phi \rangle =v,\langle \psi\rangle =0\right) = 
    \begin{pmatrix}
        h_{IJ} & 0 & 0 \\
        0 & 0 & k_{\bar r p} \\
        0 & - k_{\bar p r} & 0
    \end{pmatrix} \,,
\end{align}
where we group the scalars and fermions into the multiplet
\begin{align}
    \Phi^{a} = 
    \begin{pmatrix}
        \phi^{I} \\
        \psi^{p} \\
        \bar \psi^{\bar p}
    \end{pmatrix} \,. 
\end{align}
Lower-case Latin letters from the beginning of the alphabet run over both scalar indices $I,J,K,\dots$ and fermion indices $p,\bar p, r,\bar r,\dots$
Since the metric is evaluated at the VEV, where by assumption the fermions vanish, this metric does not contain enough information to derive all descendant geometric quantities. This is because we need to evaluate derivatives of the metric with respect to the fermion fields. However, starting from an ansatz and requiring that the metric transforms as a tensor under field redefinitions, we uniquely fix the metric. The combined scalar-fermion metric---not evaluated at the VEV---is
\begin{align}
    \bar g_{ab}(\phi,\psi) = 
    \begin{pmatrix}
        h_{IJ} & \;\;\;-\left( \tfrac{1}{2} k_{\bar sr,I} - \omega_{\bar sr I} \right) \bar \psi^{\bar s}\;\;\; & \left( \tfrac{1}{2} k_{\bar rs,I} + \omega_{\bar rs I} \right) \psi^{s}  \\
        \left( \tfrac{1}{2} k_{\bar sp,J} - \omega_{\bar sp J} \right) \bar \psi^{\bar s} & 0 & k_{\bar r p} \\
        -\left( \tfrac{1}{2} k_{\bar ps,J} + \omega_{\bar ps J} \right) \psi^{s} & - k_{\bar p r} & 0
    \end{pmatrix} \,.
\end{align}
From this metric we derive the Christoffel symbol and Riemann curvature. In contrast to the definition in \cref{eq:ChristoffelScalar} for the scalar sector, the Christoffel symbol on a supermanifold differs in that various sign factors appear in the expression, which depend on the Grassmann nature of the components. 
Instead of letting the reader combat this plethora of sign factors, we explicitly provide the various components of the Christoffel symbol, which are
\begin{align}\label{eq:ChristoffelFermion}
    \bar \Gamma^{I}_{JK} =& \Gamma^{I}_{JK} \,, \nonumber\\
    \bar \Gamma^{p}_{I s} =& \bar \Gamma^{p}_{s I} = k^{p\bar r}\left( \frac{1}{2} k_{\bar r s,I} + \omega_{\bar r s I} \right) \,, \nonumber \\
    \bar \Gamma^{\bar p}_{I \bar s} =& \bar \Gamma^{\bar p}_{\bar s I} = \left( \frac{1}{2} k_{\bar s r,I} - \omega_{\bar s r I} \right) k^{r\bar p} \,, 
\end{align}
and all other components are zero when evaluated at the VEV. The field-space covariant derivative $\bar \nabla_{a}$ uses these connections. These connections coincide with the connections in ref.~\cite{Finn:2020nvn}, although the routes to obtain them differ. 
As anticipated from the discussion about field redefinitions, both $k_{\bar pr}$ and $\omega_{\bar pr I}$ make their appearances in the connections for the field-space geometry.

The Riemann curvature is
\begin{align}\label{eq:CurvatureFermion}
    \bar R_{\bar p r IJ} &= \omega_{\bar p r  J,I} - \left(\frac{1}{2} k_{\bar ps, I} - \omega_{\bar ps I} \right) k^{s\bar t} \left(\frac{1}{2} k_{\bar tr, J} + \omega_{\bar tr J} \right) - \left( I \leftrightarrow J \right) \,,
\end{align}
again evaluated at the VEV. As expected, the Riemann curvature satisfies all the usual symmetry and Bianchi identities appropriate for a curvature on a supermanifold \cite{Arnowitt:1975bd}. For instance, the curvature with scalar indices $\bar R_{IJKL}$ is antisymmetric in $IJ$ and antisymmetric in $KL$. The mixed curvature $\bar R_{\bar p r IJ}$ in eq.~\eqref{eq:CurvatureFermion} is antisymmetric in $IJ$, but it is symmetric in $\bar p r$, because $\bar p$ and $r$ are fermionic indices and there is an additional minus sign from their exchange.

\section{Scattering amplitudes}\label{sec:Amplitudes}
The geometric quantities defined above enter in scattering amplitudes. For simplicity, we turn off the couplings to the gauge fields along with the scalar potential $V(\phi)$ and the fermion mass matrix $\cM(\phi)$ in the Lagrangian in \cref{eq:Lagr}. We are then only left with the operators in the combined scalar-fermion field-space connections in \cref{eq:ChristoffelFermion}. Similar to the scalar case, a vielbein derived from the fermion metric appears in the interpolation between the fermion field and the scattering state \cite{Cheung:2021yog}. The vielbein is the factor for each external leg that appears in the LSZ reduction formula for the $S$-matrix.
The indices of the geometric objects in the scattering amplitudes are all contracted with these vielbeins, since the scattering amplitude is defined for external states, and not external fields. For ease of notation, we use the same indices for the quantities in the scattering amplitudes and in the Lagrangian.  
For the 4-point and 5-point amplitudes below, all momenta are taken to be incoming.

The $\psi^{p} \phi^{I} \rightarrow \psi^{\bar r} \phi^{J}$ scattering amplitude is
\begin{align}
    \cA_{pI\bar rJ} = (\bar u_{\bar r} \slashed p_{I} u_{p} ) \bar R_{\bar r p JI }\,,
\end{align}
and the $\psi^{p} \phi^{I} \rightarrow \psi^{\bar r} \phi^{J} \phi^{K}$ scattering amplitude is
\begin{align}
   \cA_{pI\bar rJK} = (\bar u_{\bar r} \slashed p_{J} u_{p} ) \bar \nabla_{K}\bar R_{\bar r p IJ } + (\bar u_{\bar r} \slashed p_{K} u_{p} ) \bar \nabla_{J}\bar R_{\bar r p IK }  \,, 
\end{align}
where the curvature $\bar R$ is defined in \cref{eq:CurvatureFermion} and 
\begin{equation}   \bar\nabla_K \bar{R}_{\bar{r}pIJ}=\bar{R}_{\bar{r}pIJ,K}-\bar\Gamma^{\bar{s}}_{\bar{r}K}\bar{R}_{\bar{s}pIJ}-\bar\Gamma^s_{pK}\bar{R}_{\bar{r}sIJ}-
\bar\Gamma^L_{IK}\bar{R}_{\bar{r}pLJ}-\bar\Gamma^L_{JK}\bar{R}_{\bar{r}pIL}\,.
\end{equation}
The scattering amplitudes take remarkably compact forms when expressed in terms of the Riemann curvature and covariant derivatives of the Riemann curvature. 

Turning back on the scalar potential and fermion mass matrix, but keeping all particles massless, the scattering amplitudes become
\begin{align} 
       \cA_{pI\bar rJ}  = & (\Bar{u}_{\bar r}\slashed{p}_I u_p) \left( \bar R_{\bar r p JI} + k^{s\bar t} \left(  \frac{\cM_{\bar r s;I}\cM_{\bar t p;J}}{ s_{\bar r I} } -\frac{\cM_{\bar r s;J}\cM_{\bar t p;I}}{ s_{pI} }\right) \right) \nonumber \\
        &- (\Bar{u}_{\bar r} u_p) \left( \cM_{\bar r p;IJ} - h^{LK} \frac{\cM_{\bar r p;L} V_{;IJK}}{ s_{IJ} }  \right) \,,
\end{align}
and
%
    \begin{align}     
    \cA_{pI\bar rJK} ={} 
        &(\bar{u}_{\bar{r}}\slashed{p}_J u_p)\bar \nabla_K\bar{R}_{\bar{r}pIJ}+(\bar{u}_{\bar{r}}\slashed{p}_K u_p)\bar \nabla_J\bar{R}_{\bar{r}pIK} 
        \nonumber \\ {}&
        + k^{s \bar t} \Bigg\{\frac{\cM_{\bar t p;J}}{s_{pJ}} \bar{R}_{\bar{r} s IK}   (\bar{u}_{\bar{r}}\slashed p_K \slashed p_J u_p) +\frac{\cM_{\bar{r} s;J}}{s_{\bar{r}J}}  \bar{R}_{\bar t p KI}(\bar{u}_{\bar{r}}\slashed p_J \slashed p_K u_p) + (IJK)\Bigg\}
         \nonumber \\ {}&
         + \left\{\frac{k^{s \bar t}k^{n \bar o}}{s_{pJ}s_{\bar{r}I}} \cM_{\bar{r} n ;I}\cM_{\bar o s ;K}\cM_{\bar t p;J}
         (\bar{u}_{\bar{r}}\slashed p_I \slashed p_J u_p) + (I\leftrightarrow J \leftrightarrow K) \right\}
        \nonumber \\ {}&
        +\Bigg\{ (\bar{u}_{\bar{r}}\slashed{p}_K u_p)  \frac12 \Bigg[ -\frac{V_{;IJM}}{s_{IJ}} h^{ML} \left[ \bar{R}_{\bar{r} p LK} + k^{s \bar t} \left( \frac{\cM_{\bar{r} s;K}\cM_{\bar t p;L}}{s_{\bar{r}K}} -\frac{\cM_{\bar{r} s;L} \cM_{\bar t p ;K}}{s_{pK}}\right) \right]
        \nonumber \\ {}&
        \qquad \qquad \qquad + k^{s \bar t} \left( \frac{\cM_{\bar{r} s;K} \cM_{\bar t a ;IJ}}{s_{\bar{r}K}}-\frac{\cM_{\bar{r} s;IJ}\cM_{\bar t p;K}}{s_{pK}}\right)  \Bigg] + (I\leftrightarrow J\leftrightarrow K) \Bigg\}
        \nonumber \\ {}&
        + (\bar{u}_{\bar{r}} u_p) \Bigg\{ -\cM_{\bar{r} p ; KJI} + \frac{\cM_{\bar{r} p;L}}{s_{p\bar{r}}} h^{LM}  \bar{R}_{KJIM} (s_{IK} - s_{JI})  + \frac{\cM_{\bar{r} p;M}}{s_{p\bar{r}}}h^{ML} V_{;KJIL} 
        \nonumber \\ {}&
        \qquad \qquad\quad +\left[ \frac12 \frac{V_{;IJM}}{s_{IJ}}h^{ML} \left ( \cM_{\bar{r} p;KL} - \frac{\cM_{\bar{r} p ; N}}{s_{p\bar{r}}} h^{NO} V_{;OKL} \right)  + (I\leftrightarrow J\leftrightarrow K) \right] \Bigg\} \,.
        \end{align}
Here $s_{ab}=(p_a+p_b)^2$, where $p_a$ is the incoming momentum on the particle line with index $a$. The covariant derivatives of $V$ are $V_{;IJK}
\equiv
\nabla_K \nabla_J \nabla_I V $ and $V_{;IJKL} \equiv 
\nabla_L \nabla_K \nabla_J \nabla_I V $. $V_{;IJK}$ is completely symmetric in its indices if evaluated at an extremum of the potential, where $\nabla_I V = 0$. In the massless limit, the mass matrix $\nabla_I \nabla_J V = 0$ and $\cM_{\bar r p}=0$, and $V_{;IJKL}$ and $\cM_{\bar r p;IJ}$ are also completely symmetric.
The notation $(IJK)$ means we sum over the three cyclic permutations in $I,J,K$ whereas $(I\leftrightarrow J\leftrightarrow K)$ means we sum over the six permutations in $I,J,K$.

All couplings in the scattering amplitude are grouped in field-space covariant combinations. These results further strengthen the claim that scattering amplitudes are composed of geometric data \cite{Volkov:1970aa,Alonso:2015fsp,Helset:2020yio,Cheung:2021yog,Cheung:2022vnd,Helset:2022tlf,Helset:2022pde}.

\section{Renormalization}\label{sec:Renormalization}

The geometric formulation also simplifies the computation of the renormalization counterterms and anomalous dimensions for an EFT.
We will calculate the divergent one-loop terms from the second variation of the action by following the procedure of ref.~\cite{tHooft:1973bhk}. See also refs.~\cite{Neufeld:1998js,Henning:2016lyp,Buchalla:2017jlu,Alonso:2017tdy,Buchalla:2019wsc} for similar functional approaches. We focus on the fermionic contributions to bosonic operators, including nontrivial metric contributions. This should be combined with the analogous geometric calculations in refs.~\cite{Alonso:2016oah,Helset:2022pde} for the contributions to the bosonic operators from scalar, gauge, and mixed scalar-gauge loops.

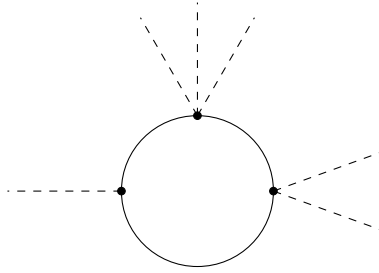
\begin{figure}
\begin{center}
\begin{tikzpicture}
\draw (0,0) circle (1);
\filldraw (0:1) circle (0.05);
\filldraw (90:1) circle (0.05);
\filldraw (180:1) circle (0.05);
\draw[dashed] (90:1) -- +(60:1.5);
\draw[dashed] (90:1) -- +(90:1.5);
\draw[dashed] (90:1) -- +(120:1.5);
\draw[dashed] (0:1) -- +(20:1.5);
\draw[dashed] (0:1) -- +(-20:1.5);
\draw[dashed] (180:1) -- +(180:1.5);
\end{tikzpicture}
\end{center}
\caption{One-loop correction to the action. The solid line is the fermion loop, and the dashed lines represent external scalar or gauge fields. There can be an arbitrary number of vertices, each of which has two fermion lines. \label{fig:loop}}
\end{figure}
The one-loop divergences are given by computing loop graphs with vertices given by the second variation of the action (see fig.~\ref{fig:loop}).
The second fermionic variation of the action is computed by writing the fermion field $\psi$ as $\psi \to \psi_B + \chi$, where $\psi_B$ is a background (external) field, and $\chi$ is a quantum field which is integrated over. The one-loop graph in fig.~\ref{fig:loop} involves the order $\chi^2$ part of the Lagrangian. The general Lagrangian bilinear in $\chi$ that we consider is
\begin{align}
    \label{eq:d2L-fermion}
	\delta_{\bar \chi \chi} S =& \int \rd^4x \ \biggl\{ \frac{1}{2} i k_{\bar p r} \left( \bar \chi^{\bar p} \gamma^{\mu}\overset{\leftrightarrow}{\mathcal{D}}_{\mu} \chi^{r} \right)
	-  \bar \chi^{\bar p} \cM_{\bar p r} \chi^{r} 
    +  \bar \chi^{\bar p} \sigma_{\mu\nu} \cT^{\mu\nu}_{\bar p r} \chi^{r} \biggr \} \,,
\end{align}
where the fermion fluctuation combines left- and right-handed degrees of freedom as 
\begin{align}\label{eq:LRfields}
    \chi = \begin{pmatrix}
        \chi_L \\
        \chi_R
    \end{pmatrix} \,.
\end{align}
The covariant derivative is $\mathcal{D}_\mu =  \partial_\mu \mathbbm{1} + \omega_\mu$.
The metric, mass, and dipole terms\footnote{We refer to $\cT^{\mu \nu}$ as the dipole term.} are written as
\begin{align} \label{eq:MTdefs}
    k =  \begin{pmatrix}
        \kappa_L & 0 \\
        0 &\kappa_R
    \end{pmatrix}\,,
    \qquad
    \cM =  \begin{pmatrix}
        0 & M \\
        M^\dag & 0
    \end{pmatrix} \,,
    \qquad
    \cT^{\mu\nu} =  \begin{pmatrix}
        0 & T^{\mu\nu} \\
        T^{\mu\nu\dag}  & 0
    \end{pmatrix} \,,
\end{align}
which are functions of the external fields (possibly including derivatives). We do not include fermion bilinears with additional derivatives acting on the fermion fields in \cref{eq:d2L-fermion}. Such terms are not needed for analyzing the original Lagrangian in \cref{eq:Lagr}.
The dipole $T^{\mu\nu}$ can be taken to be selfdual; several relations arising from this condition are given in \cref{sec:self-dual}, and were used to simplify the results.
In cases where Majorana mass or Majorana dipole terms are present, as in the LEFT~\cite{Jenkins:2017dyc,Jenkins:2017jig}, one can use the more general form for the matrices presented in appendix~\ref{app:Majorana}. Previous calculations~\cite{tHooft:1973bhk} did not include dipole couplings, which are present in a general EFT, and are needed for both the SMEFT and the LEFT.

The infinite bosonic part of the one-loop functional integral in $4-2\epsilon$ dimensions is\footnote{The counterterm Lagrangian is the negative of this expression.}
\begin{align} 
\begin{aligned}\label{eq:oneloopdivcompact}
	\Delta S =& \frac{1}{32 \pi^2 \epsilon}   \int \rd^4x \ \bigg\{ \frac{1}{3} {\rm Tr} \left[ \cY_{\mu\nu} \cY^{\mu\nu} \right] + {\rm Tr} \left[ (\cD_{\mu} \cM) (\cD^{\mu} \cM) - (\cM\cM)^2 \right] \\
 & \qquad \qquad\qquad -\frac{16}{3} {\rm Tr}[(\cD_\mu \mathcal T^{\mu \alpha})(\cD_\nu \mathcal T^{\nu \alpha})  - (\mathcal T^{\mu\nu} \mathcal T^{\alpha \beta})^2 ]\\
& \qquad \qquad\qquad -4 i {\rm Tr}[ \cY_{\mu\nu} (\mathcal M \mathcal{T^{\mu\nu}}+ \mathcal T^{\mu\nu}\mathcal M)] - 8 {\rm Tr} (\mathcal M \mathcal T^{\mu\nu})^2
\bigg\} \,,
\end{aligned}
\end{align}
where
\begin{align}
	\left[ \cY_{\mu\nu} \right]^{p}_{\;\; r} =& \left[ \cD_{\mu} , \cD_{\nu} \right]^{p}_{\;\; r} 
	= \bar R^{p}_{\;\;r IJ}(D_{\mu}\phi)^{I} (D_{\nu}\phi)^{J} + \left(\bar \nabla_{r} t^{p}_{A}\right) F^{A}_{\mu\nu}  \,, \\
	(\cD_{\mu} \cM)^{p}_{\;\; r} =& k^{p\bar t} (\cD_{\mu} \cM_{\bar t r}) = k^{p\bar t} \left[ D_{\mu} \cM_{\bar t r} - \bar \Gamma^{\bar s}_{I\bar t} (D_{\mu}\phi)^{I} \cM_{\bar s r} - \bar \Gamma^{s}_{I r} (D_{\mu}\phi)^{I} \cM_{\bar t s}  \right] \,, \\
	(\cM \cM)^{p}_{\;\; r} =& k^{p \bar t} \cM_{\bar t q} k^{q \bar s} \cM_{\bar s r} \,, \\
    (\cD_{\mu} \cT^{\alpha\beta})^{p}_{\;\; r} =& k^{p\bar t} (\cD_{\mu} \cT^{\alpha\beta}_{\bar t r}) = k^{p\bar t} \left[ D_{\mu} \cT^{\alpha\beta}_{\bar t r} - \bar \Gamma^{\bar s}_{I\bar t} (D_{\mu}\phi)^{I} \cT^{\alpha\beta}_{\bar s r} - \bar \Gamma^{s}_{I r} (D_{\mu}\phi)^{I} \cT^{\alpha\beta}_{\bar t s}  \right] \,, \\
	(\cT^{\mu\nu} \cT^{\alpha\beta})^{p}_{\;\; r} =& k^{p \bar t} \cT^{\mu\nu}_{\bar t q} k^{q \bar s} \cT^{\alpha\beta}_{\bar s r}  \,.
\end{align}
The connections and curvature are defined in \cref{eq:ChristoffelFermion,eq:CurvatureFermion}, while the covariant derivative of the Killing vector is $\bar \nabla_{r} t^{p}_{A} = t^{p}_{A,r} + \bar \Gamma^{p}_{Ir} t^{I}_{A}$. The final expression involves the field-strength tensor computed from the commutator of two fermion covariant derivatives, as well as covariant derivatives of the mass and dipole matrices. The field-strength tensor $ \cY_{\mu\nu}$ has the same formula as for the scalar case~\cite{Alonso:2016oah} and the scalar-gauge case~\cite{Helset:2022pde}, with the curvature tensor now having two fermionic and two bosonic indices, rather than four bosonic indices.
From the above expressions we can calculate the one-loop anomalous dimensions of the effective operators in the Lagrangian. In the next section we apply the general formula in \cref{eq:oneloopdivcompact} to the SMEFT, for both dimension-six and dimension-eight operators.

\section{Standard Model Effective Field Theory}\label{sec:smeft}

The results in the previous sections hold for a general effective field theory. In this section, we will apply these results to the SMEFT. 

The Standard Model Lagrangian is
\begin{align}
	\cL = - \frac{1}{4} F^{A}_{\mu\nu} F^{A\,\mu\nu} + (D_{\mu} H)^{\dagger} (D^{\mu} H) - \lambda \left( H^{\dagger} H - \frac{1}{2} v^2 \right)^{2}
	+ \delta_{\bar p r} i \bar \psi^{\bar p} \gamma^{\mu} D_{\mu} \psi^{r} - \bar \psi^{\bar p} \cM_{{\rm SM,}\bar p r} \psi^{r} \,,
\end{align}
where we have grouped all gauge fields from the gauge group $SU(3)_{c} \otimes SU(2)_{L} \otimes U(1)_{Y}$ into the multiplet
\begin{align}
	A^{B}_{\mu} = \begin{pmatrix}
		G^{\mathscr{B}}_{\mu} \\
		W^{b}_{\mu} \\
		B_{\mu}
	\end{pmatrix} , \qquad F^{B}_{\mu\nu} = \partial_{\mu} A^{B}_{\nu} - \partial_{\nu} A^{B}_{\mu} - f^{B}_{\;\;CD} A^{C}_{\mu} A^{D}_{\nu} \,.
\end{align}
Similarly, all fermions are grouped into the multiplet
\begin{align}
	\psi^{p} = 
	\begin{pmatrix}
		\ell^{p}_{L} \\ 
		q^{p}_{L} \\
		e^{p}_{R} \\
		u^{p}_{R} \\
		d^{p}_{R} 
	\end{pmatrix} \,,
\end{align}
where the index $p$ runs over all flavor indices and gauge group indices. The Yukawa terms are all grouped into one term,
\begin{align}
	\bar \psi^{\bar p} \cM_{{\rm SM,}\bar pr} \psi^{r} = [Y_{e}]^{\dagger}_{\bar pr} (\bar L_{\bar p} e_{r} H) +  [Y_{u}]^{\dagger}_{\bar pr} (\bar Q_{\bar p} u_{r} \tilde H) +  [Y_{d}]^{\dagger}_{\bar pr} (\bar Q_{\bar p} d_{r} H) + {\rm h.c.} 
\end{align}
The fermion metric and dipole are trivial in the Standard Model,
\begin{align}
    k_{{\rm SM},\bar pr} =~& \delta_{\bar pr} \,, &
    \cT^{\mu\nu}_{\rm SM} =~& 0 \,. 
\end{align}

The SMEFT extends the Standard Model by adding a complete, non-redundant set of higher-dimensional operators built from the fields of the Standard Model. For our purposes, we focus on the operators with fermions at mass dimension six and eight that contribute to \cref{eq:Lagr}. These operators are listed in \cref{dim:6_fermion} and \cref{dim:8_fermion}, respectively. The dimension-six operators are all in the Warsaw basis \cite{Grzadkowski:2010es}, and the dimension-eight operators coincide with the operator basis in ref.~\cite{Murphy:2020rsh},\footnote{An operator basis at mass dimension eight was also constructed in ref.~\cite{Li:2020gnx}.} with one exception: the operators $\op{8}{l^2 H^4 D}{(2)}$ and $\op{8}{l^2 H^4 D}{(3)}$ in \cref{dim:8_fermion} are linear combinations of two operators in ref.~\cite{Murphy:2020rsh}, and similarly for the corresponding operators with quarks. 

We will illustrate how the operators in the SMEFT can be grouped into the scalar and dipole terms previously defined. We will focus on the terms involving the right-handed electron fields for simplicity; all other components are extracted analogously. The mass matrix is
\begin{align}
    M_{\bar pr} \supset [Y_{e}]^\dag_{\bar pr} H - \coef{6}{\underset{\bar pr}{leH^3}}{} H (H^\dag H) - \coef{8}{\underset{\bar pr}{leH^5}}{} H (H^\dag H)^2 \,,
\end{align}
where in addition to the Yukawa couplings, we also have dimension-six and dimension-eight contributions. We have included a left superscript 6 or 8 to make clear which contributions are dimension six or dimension eight.
This coincides with the grouping of terms introduced in ref.~\cite{Helset:2020yio}. Similarly, the dipole term with the $B_{\mu\nu}$ field strength up to mass dimension eight is
\begin{align}
    T^{\mu\nu}_{\bar pr} \supset \coef{6}{\underset{\bar pr}{leBH}}{} H \frac{1}{2}\left(B^{\mu\nu} - i \tilde B^{\mu\nu}\right) + \coef{8}{\underset{\bar pr}{leBH^3}}{} H (H^{\dagger} H) \frac{1}{2}\left(B^{\mu\nu} - i \tilde B^{\mu\nu}\right) \,. 
\end{align}
Note that the dipole is explicitly selfdual; this is required to use eq.~\eqref{eq:oneloopdivcompact}.

We apply the formalism developed above to the SMEFT. To illustrate the method, we include SMEFT operators that contribute to the mass, dipole, and fermion covariant derivative. The SMEFT dimension-six operators of type $\psi^2 H^3$ and dimension-eight operators of type $\psi^2 H^5$ are included in the mass matrix,  dimension-six operators of type $\psi^2 XH$ and dimension-eight operators of type $\psi^2 X H^3$ in the dipole matrix, and  dimension-six operators of type $\psi^ 2H^2 D$ and dimension-eight operators of type $\psi^2 H^4 D$ in the fermion covariant derivative.

There are additional dimension-eight SMEFT operators which can also be included --- $\psi^2 X^2 H$ and $\psi^2 H^3 D^2$ with no fermion derivatives could be included in the mass, $\psi^2 X^2 H$, $\psi^2 XH D^2$ with no fermion derivatives, and $\psi^2 H^3 D^2$ could be included in the dipole, and $\bar R R X H^2 D$ and $\bar LL X H^2 D$ could be added to the fermion covariant derivative. These do not introduce any new features in the calculation. However, they complicate the final RGE, and so will be discussed elsewhere.

Four-fermion operators can also be included in our analysis using Fierz identities. In a four-fermion operator, two fermion fields are treated as background fields, and two as quantum fields. Writing $\psi = \psi_B +\chi$ in terms of background and quantum fields, we get
\begin{align}
( \bar \psi_1 \Gamma \psi_2 )( \bar \psi_3 \Gamma \psi_4 ) &\to  ( \bar \psi_{1,B} \Gamma \psi_{2,B} )( \bar \chi_3 \Gamma \chi_4 )
 + ( \bar \psi_{1,B} \Gamma \chi_2 )( \bar \psi_{3,B} \Gamma \chi_4 )
 + ( \bar \psi_{1,B} \Gamma \chi_2 )( \bar \chi_3 \Gamma \psi_{4,B} ) \nonumber \\ 
& +  ( \bar \chi_1 \Gamma \psi_{2,B} )( \psi_{3,B} \Gamma \chi_4 )
 +  ( \bar \chi_1 \Gamma \psi_{2,B} )( \chi_3 \Gamma \psi_{4,B} )
 +  ( \bar \chi_1 \Gamma \chi_2 )( \psi_{3,B} \Gamma \psi_{4,B} )\,.
\end{align}
The first and last terms on the right-hand side are included in the quadratic fermion Lagrangian in eq.~\eqref{eq:d2L-fermion} in either the mass, tensor, or covariant derivative depending on whether $\Gamma=1,\sigma^{\mu\nu},\gamma^\mu$. The coefficient $( \bar \psi_{1,B} \Gamma \psi_{2,B} )$ is an external bosonic source, since it contains two fermion fields. The other terms can be treated the same way, by Fierzing so that the two quantum fields are in a single fermion bilinear.\footnote{
Using Fierz identities is sufficient for tree-level amplitudes and one-loop renormalization. However, since the Fierz identities are valid only in four dimension, we need a different approach to describe the one-loop finite terms and higher-loop effects from four-fermion operators.
} These terms will also be discussed elsewhere.

We proceed to calculate the renormalization group equations for bosonic operators coming from a fermion loop in the SMEFT to mass dimension eight. This means that we can directly apply the general formula in \cref{eq:oneloopdivcompact} to the SMEFT operators. As in ref.~\cite{Helset:2022pde}, evaluating the expressions in \cref{eq:oneloopdivcompact} leads to operators which are not necessarily in the chosen operator basis. They can be converted to the standard basis by integration-by-parts relations and by field redefinitions.
Many of these relations were worked out in ref.~\cite{Helset:2022pde}, so we can reuse them here. When the smoke clears, we are only left with operators in our basis, and we can read off the RGE from their coefficients. The renormalization group equations for bosonic SMEFT operators from fermionic operators to mass dimension eight are listed in appendix~\ref{sec:RGEresults}.

\section{Low-energy Effective Field Theory}\label{sec:LEFT}

The effective field theory below the electroweak scale is the Low-energy Effective Field Theory, or LEFT, which is obtained from the SMEFT after integrating out all fields with masses on the order of the electroweak scale \cite{Fermi:1934sk,Jenkins:2017jig,Jenkins:2017dyc,Dekens:2019ept}. This includes the Higgs field. Since the low-energy theory has no scalar fields, many of the previous geometric considerations are less important. For instance, the metric is trivial,
\begin{align}
    k_{{\rm LEFT},\bar pr} = \delta_{\bar pr} \,. 
\end{align}
However, we can still apply the general formula in \cref{eq:oneloopdivcompact} to derive the RGE coming from double insertions of dipole operators, $T^{\mu\nu}$, which are dimension five in the LEFT. This serves as an independent cross-check on parts of the RGE results in ref.~\cite{Jenkins:2017dyc}. The relevant dipole operators in the LEFT are
\begin{align}
    \sum_{\psi=e,u,d} \left( L_{\underset{rs}{\psi\gamma}} (\bar e_{Lp} \sigma^{\mu\nu} e_{Rr}) F_{\mu\nu} + {\rm h.c.} \right) + \left( L_{\underset{rs}{\nu\gamma}}(\nu^{T}_{Lp}C \sigma^{\mu\nu} \nu_{Lr})F_{\mu\nu} + {\rm h.c.} \right) \,,
\end{align}
which contribute to the RGE for the four-fermion operator
\begin{align}
    L^{V,LL}_{\underset{prst}{ee}} (\bar e_{Lp} \gamma^{\mu} e_{Lr}) (\bar e_{Ls} \gamma_{\mu} e_{Lt}) \,.
\end{align}
We obtain using \cref{eq:oneloopdivcompact} that
\begin{align}
    \dot L^{V,LL}_{\underset{prst}{ee}} \supset e^2 q^2_{e} \delta_{pr}\delta_{st} \zeta_{e} \,, 
\end{align}
where
\begin{align}
    \zeta_{e} = \frac{8}{3} {\rm Tr}\left[2 L_{\nu\gamma} L^{\dagger}_{\nu\gamma} + L_{e\gamma} L^{\dagger}_{e\gamma} + N_{c} L_{u\gamma} L^{\dagger}_{u\gamma} + N_{c} L_{d\gamma} L^{\dagger}_{d\gamma}  \right] \,. 
\end{align}
These terms arise from the quadratic terms in $T^{\mu\nu}$ in \cref{eq:oneloopdivcompact}, after applying a field redefinition. We also find similar results for the RGE of other four-fermion operators in the LEFT, in agreement with ref.~\cite{Jenkins:2017dyc}.

\section{Conclusion}\label{sec:Conclusion}

We have formulated a field-space geometry for scalars and fermions. This geometric picture simplifies calculations and makes the covariance of scattering amplitudes under field redefinitions manifest by expressing the scattering of particles in terms of the Riemann curvature in field space and covariant derivatives.  We then used the geometry to calculate the fermionic contributions to the bosonic RGEs in the SMEFT to mass dimension eight. The results are listed in app.~\ref{sec:RGEresults}. 

The RGEs for fermionic operators were not the focus in this work. However, there is no obstacle in computing them using the geometric tools developed in this paper. We will return to the evaluation of these RGEs in the future.

The field-space geometry considered in this paper is linked to field redefinitions with no derivatives. Scattering amplitudes are insensitive to a larger class of field redefinitions including derivatives. Some work has been done to incorporate these general field redefinitions into the geometric language \cite{Cheung:2022vnd,Cohen:2022uuw,Craig:2023wni}. However, a complete description of the field-space geometry with fermions and higher-derivative field redefinitions is still lacking. We leave this for future work.

\subsection*{Acknowledgments}

We thank Xiaochuan Lu for helpful discussions.
This work is supported in part by the U.S.\ Department of Energy (DOE) under award numbers~DE-SC0009919 and DE-SC0011632 and by the Walter Burke Institute for Theoretical Physics. This work is also supported in part by the Fermi National Accelerator Laboratory (Fermilab), a U.S. Department of Energy, Office of Science, HEP User Facility.
Fermilab is managed by Fermi Research Alliance, LLC (FRA), acting under Contract No. DE--AC02--07CH11359.

\appendix

\section{Operators}\label{sec:Operators}

Here we list the operator basis used in this work. The bosonic SMEFT operators at dimension six are in Table~\ref{dim:6}, while the dimension-six operators with a fermion bilinear are given in Table~\ref{dim:6_fermion}. The Warsaw basis \cite{Grzadkowski:2010es} also contains four-fermion operators, but we do not include them in this work. The SMEFT operators at dimension eight are similarly split into bosonic operators in Table~\ref{dim:8} and fermion bilinear operators in Table~\ref{dim:8_fermion}. Our convention for these operators aligns with the operator basis in ref.~\cite{Murphy:2020rsh}, with a few exceptions noted in the main text.
%
\begin{table}
\begin{center}
\begin{minipage}[t]{5cm}
\vspace{-0.5cm}
\renewcommand{\arraystretch}{1.5}
\begin{align*}
\begin{array}{c|c|c}
\multicolumn{3}{c}{\bm{X^3}} \\
\hline
Q_G     & \op{6}{G^3}{}           & f^{\mathscr{A}\mathscr{B}\mathscr{C}} G_\mu^{\mathscr{A}\nu} G_\nu^{\mathscr{B}\rho} G_\rho^{\mathscr{C}\mu}  \\
Q_W      & \op{6}{W^3}{}            & \epsilon^{abc} W_\mu^{a\,\nu} W_\nu^{b\,\rho} W_\rho^{c \, \mu} \\ 
\end{array}
\end{align*}
\renewcommand{\arraystretch}{1.5}
\begin{align*}
\begin{array}{c|c|c}
\multicolumn{3}{c}{\bm{H^4 D^2} } \\
\hline
Q_{H\Box} & \op{6}{H^4\Box}{} & (H^\dag H)\Box (H^\dag H) \\
Q_{H D} &\op{6}{H^4D^2}{}  & \left(D^\mu H^\dag H\right) \left(H^\dag D_\mu H\right)
\end{array}
\end{align*}
\vspace{-0.5cm}
\renewcommand{\arraystretch}{1.5}
\begin{align*}
\begin{array}{c|c|c}
\multicolumn{3}{c}{\bm{H^6} } \\
\hline
Q_H    & \op{6}{H^6}{}   & (H^\dag H)^3
\end{array}
\end{align*}
\end{minipage}
\hspace{1cm}
\begin{minipage}[t]{2.5cm}
\vspace{-0.5cm}
\renewcommand{\arraystretch}{1.5}
\begin{align*}
\begin{array}{c|c|c}
\multicolumn{3}{c}{\bm{X^2 H^2} } \\
\hline
Q_{H G}  & \op{6}{G^2H^2}{(1)}   & (H^\dag H)\, G^{\mathscr{A}}_{\mu\nu} G^{\mathscr{A}\mu\nu} \\
Q_{H \tilde G}  & \op{6}{G^2H^2}{(2)}   & (H^\dag H)\, G^{\mathscr{A}}_{\mu\nu} \tilde G^{\mathscr{A}\mu\nu} \\
Q_{H W}  &\op{6}{W^2H^2}{(1)}     & (H^\dag H) \, W^a_{\mu\nu} W^{a\mu\nu} \\
Q_{H \tilde W}  &\op{6}{W^2H^2}{(2)}     & (H^\dag H) \, W^a_{\mu\nu} \tilde W^{a\mu\nu} \\
Q_{H B} & \op{6}{B^2H^2}{(1)}      &  (H^\dag H)\, B_{\mu\nu} B^{\mu\nu} \\
Q_{H \tilde B} & \op{6}{B^2H^2}{(2)}      &  (H^\dag H)\, B_{\mu\nu} \tilde B^{\mu\nu} \\
Q_{H WB} &  \op{6}{WBH^2}{(1)}    &  (H^\dag \tau^a H)\, W^a_{\mu\nu} B^{\mu\nu} \\
Q_{H \tilde WB} &  \op{6}{WBH^2}{(2)}    &  (H^\dag \tau^a H)\, \tilde W^a_{\mu\nu} B^{\mu\nu} \\
\end{array}
\end{align*}
\end{minipage}
\end{center}
\caption{\label{dim:6} Bosonic dimension-six operators in the SMEFT. The first column is the notation of ref.~\cite{Grzadkowski:2010es}, and the second column is the notation used in this paper.}
\end{table}
%

%
\begin{table}
\begin{center}
\begin{minipage}[t]{5cm}
\vspace{-0.5cm}
\renewcommand{\arraystretch}{1.5}
\begin{align*}
\begin{array}{c|c|c}
\multicolumn{3}{c}{\bm{\psi^2 H^3 + \textbf{h.c.}}} \\
\hline
Q_{eH}    & \op{6}{\ell e H^3}{}           & (H^{\dagger} H)(\bar \ell_{p} e_{r} H)  \\
Q_{uH}    & \op{6}{q u H^3}{}           & (H^{\dagger} H)(\bar q_{p} u_{r}\tilde H)  \\
Q_{dH}    & \op{6}{q d H^3}{}           & (H^{\dagger} H)(\bar q_{p} d_{r} H)  \\
\end{array}
\end{align*}
\renewcommand{\arraystretch}{1.5}
\begin{align*}
\begin{array}{c|c|c}
\multicolumn{3}{c}{\bm{\psi^2 X H + \textbf{h.c.}}} \\
\hline
Q_{eW} & \op{6}{\ell e W H}{} & (\bar \ell_{p} \sigma^{\mu\nu} \tau^{a} e_{r} H) W^{a}_{\mu\nu} \\
Q_{eB} & \op{6}{\ell e B H}{} & (\bar \ell_{p} \sigma^{\mu\nu}  e_{r} H) B_{\mu\nu} \\
Q_{uG} & \op{6}{q u G H}{} & (\bar q_{p} \sigma^{\mu\nu} T^{\mathscr{A}} u_{r} \tilde H) G^{\mathscr{A}}_{\mu\nu} \\
Q_{uW} & \op{6}{q u W H}{} & (\bar q_{p} \sigma^{\mu\nu} \tau^{a} u_{r} \tilde H) W^{a}_{\mu\nu} \\
Q_{uB} & \op{6}{q u B H}{} & (\bar q_{p} \sigma^{\mu\nu} u_{r} \tilde H) B_{\mu\nu} \\
Q_{dG} & \op{6}{q d G H}{} & (\bar q_{p} \sigma^{\mu\nu} T^{\mathscr{A}} d_{r} H) G^{\mathscr{A}}_{\mu\nu} \\
Q_{dW} & \op{6}{q d W H}{} & (\bar q_{p} \sigma^{\mu\nu} \tau^{a} d_{r} H) W^{a}_{\mu\nu} \\
Q_{dB} & \op{6}{q d B H}{} & (\bar q_{p} \sigma^{\mu\nu} d_{r} H) B_{\mu\nu} \\
\end{array}
\end{align*}
\end{minipage}
\hspace{1cm}
\begin{minipage}[t]{2.5cm}
\vspace{-0.5cm}
\renewcommand{\arraystretch}{1.5}
\renewcommand{\arraystretch}{1.5}
\begin{align*}
\begin{array}{c|c|c}
\multicolumn{3}{c}{\bm{\psi^2 H^2 D} } \\
\hline
Q_{H \ell}^{(1)}  & \op{6}{\ell^2H^2 D}{(1)}   & (\bar \ell_{p} \gamma^{\mu} \ell_{r})\,(H^\dag i \overset{\leftrightarrow}{D}_{\mu} H)  \\
Q_{H \ell}^{(3)}  & \op{6}{\ell^2H^2 D}{(3)}   & (\bar \ell_{p} \gamma^{\mu} \tau^{a} \ell_{r})\,(H^\dag i \overset{\leftrightarrow}{D}_{\mu}^{a} H)  \\
Q_{H e}  & \op{6}{e^2H^2 D}{}   & (\bar e_{p} \gamma^{\mu} e_{r})\,(H^\dag i \overset{\leftrightarrow}{D}_{\mu} H)  \\
Q_{H q}^{(1)}  & \op{6}{q^2H^2 D}{(1)}   & (\bar q_{p} \gamma^{\mu} q_{r})\,(H^\dag i \overset{\leftrightarrow}{D}_{\mu} H)  \\
Q_{H q}^{(3)}  & \op{6}{q^2H^2 D}{(3)}   & (\bar q_{p} \gamma^{\mu}\tau^a q_{r})\,(H^\dag i \overset{\leftrightarrow}{D}_{\mu}^{a} H)  \\
Q_{H u}  & \op{6}{u^2H^2 D}{}   & (\bar u_{p} \gamma^{\mu} u_{r})\,(H^\dag i \overset{\leftrightarrow}{D}_{\mu} H)  \\
Q_{H d}  & \op{6}{d^2H^2 D}{}   & (\bar d_{p} \gamma^{\mu} d_{r})\,(H^\dag i \overset{\leftrightarrow}{D}_{\mu} H)  \\
Q_{H ud}  & \op{6}{udH^2 D}{}   & (\bar u_{p} \gamma^{\mu} d_{r})\, (\tilde H^\dag i D_{\mu} H) + \textrm{h.c.}  \\
\end{array}
\end{align*}
\end{minipage}
\end{center}
\caption{\label{dim:6_fermion} Fermionic dimension-six operators in the SMEFT (not including four-fermion operators). The first column is the notation of ref.~\cite{Grzadkowski:2010es}, and the second column is the notation used in this paper.}
\end{table}
%

\begin{table}
\begin{center}
\begin{minipage}[t]{2.15cm}
\vspace{-0.5cm}
\renewcommand{\arraystretch}{1.5}
\begin{align*}
\begin{array}{c|c}
\multicolumn{2}{c}{\bm{H^8} } \\
\hline
\op{8}{H^8}{} &  (H^\dag H)^4 
\end{array}
\end{align*}
\vspace{-0.5cm}
\renewcommand{\arraystretch}{1.5}
\begin{align*}
\begin{array}{c|c}
\multicolumn{2}{c}{\bm{H^6 D^2} } \\
\hline
\op{8}{H^6 D^2}{(1)}  & (H^{\dag} H)^2 (D_{\mu} H^{\dag} D^{\mu} H) \\
\op{8}{H^6 D^2}{(2)}  & (H^{\dag} H) (H^{\dag} \tau^I H) (D_{\mu} H^{\dag} \tau^I D^{\mu} H)
\end{array}
\end{align*}

\vspace{-0.5cm}
\renewcommand{\arraystretch}{1.5}
\begin{align*}
\begin{array}{c|c}
\multicolumn{2}{c}{\bm{H^4 D^4} } \\
\hline
\op{8}{H^4 D^4}{(1)}  &  (D_{\mu} H^{\dag} D_{\nu} H) (D^{\nu} H^{\dag} D^{\mu} H) \\ 
\op{8}{H^4 D^4}{(2)}  &  (D_{\mu} H^{\dag} D_{\nu} H) (D^{\mu} H^{\dag} D^{\nu} H) \\ 
\op{8}{H^4 D^4}{(3)}  &  (D^{\mu} H^{\dag} D_{\mu} H) (D^{\nu} H^{\dag} D_{\nu} H)
\end{array}
\end{align*}
\vspace{-0.5cm}
\renewcommand{\arraystretch}{1.5}
\begin{align*}
\begin{array}{c|c}
\multicolumn{2}{c}{\bm{X^3 H^2} } \\
\hline
\op{8}{G^3H^2}{(1)}  &  f^{\mathscr{A}\mathscr{B}\mathscr{C}} (H^\dag H) G_{\mu}^{\mathscr{A}\nu} G_{\nu}^{\mathscr{B}\rho} G_{\rho}^{\mathscr{C}\mu} \\
\op{8}{W^3H^2}{(1)}  &  \epsilon^{abc} (H^\dag H) W_{\mu}^{a\,\nu} W_{\nu}^{b\,\rho} W_{\rho}^{c\,\mu} \\
\op{8}{W^2BH^2}{(1)}  &  \epsilon^{abc} (H^\dag \tau^a H) B_{\mu}^{\;\;\,\nu} W_{\nu}^{b\,\rho} W_{\rho}^{c\,\mu} \\
\end{array}
\end{align*}
\vspace{-0.5cm}
\renewcommand{\arraystretch}{1.5}
\begin{align*}
\begin{array}{c|c}
\multicolumn{2}{c}{\bm{X^2 H^4} } \\
\hline
\op{8}{G^2H^4}{(1)}  & (H^\dag H)^2 G^{\mathscr{A}}_{\mu\nu} G^{\mathscr{A}\mu\nu} \\
\op{8}{G^2H^4}{(2)}  & (H^\dag H)^2 G^{\mathscr{A}}_{\mu\nu} \tilde G^{\mathscr{A}\mu\nu} \\
\op{8}{W^2H^4}{(1)}  & (H^\dag H)^2 W^a_{\mu\nu} W^{a\,\mu\nu} \\
\op{8}{W^2H^4}{(2)}  & (H^\dag H)^2 W^a_{\mu\nu} \tilde W^{a\,\mu\nu} \\
\op{8}{W^2H^4}{(3)}  & (H^\dag \tau^a H) (H^\dag \tau^b H) W^a_{\mu\nu} W^{b\,\mu\nu} \\
\op{8}{W^2H^4}{(4)}  & (H^\dag \tau^a H) (H^\dag \tau^b H) W^a_{\mu\nu} \tilde W^{b\,\mu\nu} \\
\op{8}{WBH^4}{(1)}  &  (H^\dag H) (H^\dag \tau^a H) W^a_{\mu\nu} B^{\mu\nu} \\
\op{8}{WBH^4}{(2)}  &  (H^\dag H) (H^\dag \tau^a H) \tilde W^a_{\mu\nu} B^{\mu\nu} \\
\op{8}{B^2H^4}{(1)}  &  (H^\dag H)^2 B_{\mu\nu} B^{\mu\nu} \\
\op{8}{B^2H^4}{(2)}  &  (H^\dag H)^2 B_{\mu\nu} \tilde B^{\mu\nu} \\
\end{array}
\end{align*}
\end{minipage}
\hspace{0.5cm}
\begin{minipage}[t]{4cm}
\vspace{-0.5cm}
\renewcommand{\arraystretch}{1.5}
\begin{align*}
\begin{array}{c|c}
\multicolumn{2}{c}{\bm{X  H^4 D^2} } \\
\hline
\op{8}{WH^4D^2}{(1)}  & i (H^{\dag} H) (D^{\mu} H^{\dag} \tau^a D^{\nu} H) W_{\mu\nu}^a \\
\op{8}{WH^4D^2}{(2)}  & i (H^{\dag} H) (D^{\mu} H^{\dag} \tau^a D^{\nu} H) \tilde W_{\mu\nu}^a \\
\op{8}{WH^4D^2}{(3)}  & i \epsilon^{abc} (H^{\dag} \tau^a H) (D^{\mu} H^{\dag} \tau^b D^{\nu} H) W_{\mu\nu}^c \\
\op{8}{WH^4D^2}{(4)}  & i \epsilon^{abc} (H^{\dag} \tau^a H) (D^{\mu} H^{\dag} \tau^b D^{\nu} H) \tilde W_{\mu\nu}^c \\
\op{8}{BH^4D^2}{(1)}  & i (H^{\dag} H) (D^{\mu} H^{\dag} D^{\nu} H) B_{\mu\nu} \\
\op{8}{BH^4D^2}{(2)}  & i (H^{\dag} H) (D^{\mu} H^{\dag} D^{\nu} H) \tilde B_{\mu\nu} \\
\end{array}
\end{align*}

\vspace{-0.5cm}
\renewcommand{\arraystretch}{1.5}
\begin{align*}
\begin{array}{c|c}
\multicolumn{2}{c}{\bm{X^2 H^2 D^2} } \\
\hline
\op{8}{G^2H^2D^2}{(1)}  &  (D^{\mu} H^{\dag} D^{\nu} H) G_{\mu\rho}^{\mathscr{A}} G_{\nu}^{\mathscr{A} \rho} \\
\op{8}{G^2H^2D^2}{(2)}  &  (D^{\mu} H^{\dag} D_{\mu} H) G_{\nu\rho}^{\mathscr{A}} G^{\mathscr{A} \nu\rho} \\
\op{8}{W^2H^2D^2}{(1)}  &  (D^{\mu} H^{\dag} D^{\nu} H) W_{\mu\rho}^a W_{\nu}^{a\, \rho} \\
\op{8}{W^2H^2D^2}{(2)}  &  (D^{\mu} H^{\dag} D_{\mu} H) W_{\nu\rho}^a W^{a\, \nu\rho} \\
\op{8}{W^2H^2D^2}{(4)}  &  i \epsilon^{abc} (D^{\mu} H^{\dag} \tau^a D^{\nu} H) W_{\mu\rho}^b W_{\nu}^{c\, \rho} \\
\op{8}{W^2H^2D^2}{(5)}  &  \epsilon^{abc} (D^{\mu} H^{\dag} \tau^a D^{\nu} H) \\  & \times (W_{\mu\rho}^b \tilde W_{\nu}^{c\, \rho} - \tilde W_{\mu\rho}^b W_{\nu}^{c\, \rho}) \\
\op{8}{WBH^2D^2}{(1)}  &  (D^{\mu} H^{\dag} \tau^a D_{\mu} H) B_{\nu\rho} W^{a\, \nu\rho} \\
\op{8}{WBH^2D^2}{(2)}  &  (D^{\mu} H^{\dag} \tau^a D_{\mu} H) B_{\nu\rho} \tilde W^{a\, \nu\rho} \\
\op{8}{WBH^2D^2}{(3)}  &  i (D^{\mu} H^{\dag} \tau^a D^{\nu} H) \\
 &  \times (B_{\mu\rho} W_{\nu}^{a\, \rho} - B_{\nu\rho} W_{\mu}^{a\,\rho}) \\
\op{8}{WBH^2D^2}{(4)}  &  (D^{\mu} H^{\dag} \tau^a D^{\nu} H) \\
  & \times (B_{\mu\rho} W_{\nu}^{a\, \rho} + B_{\nu\rho} W_{\mu}^{a\,\rho}) \\
\op{8}{WBH^2D^2}{(6)}  &  (D^{\mu} H^{\dag} \tau^a D^{\nu} H) \\
  & \times (B_{\mu\rho} \tilde W_{\nu}^{a\, \rho} + B_{\nu\rho} \tilde W_{\mu}^{a\,\rho}) \\
\op{8}{B^2H^2D^2}{(1)}  &  (D^{\mu} H^{\dag} D^{\nu} H) B_{\mu\rho} B_{\nu}^{\,\,\,\rho} \\
\op{8}{B^2H^2D^2}{(2)}  &  (D^{\mu} H^{\dag} D_{\mu} H) B_{\nu\rho} B^{\nu\rho} \\
\end{array}
\end{align*}

\end{minipage}
\hspace{-1.5cm}
%
\end{center}
\caption{\label{dim:8} Bosonic dimension-eight operators in the SMEFT.  The $XH^4D^2$ operators have a factor of $i$ relative to ref.~\cite{Murphy:2020rsh} to make them hermitian. There are also $X^4$ operators which have not been listed.}
\end{table}

%
\begin{table}
\begin{center}
\begin{minipage}[t]{5cm}
\vspace{-0.5cm}
\renewcommand{\arraystretch}{1.5}
\begin{align*}
\begin{array}{c|c}
\multicolumn{2}{c}{\bm{\psi^2 H^5 + \textbf{h.c.}}} \\
\hline
  \op{8}{\ell e H^5}{}           & (H^{\dagger} H)^{2} (\bar \ell_{p} e_{r} H)  \\
  \op{8}{q u H^5}{}           & (H^{\dagger} H)^{2} (\bar q_{p} u_{r}\tilde H)  \\
  \op{8}{q d H^5}{}           & (H^{\dagger} H)^{2} (\bar q_{p} d_{r} H)  \\
\end{array}
\end{align*}
\renewcommand{\arraystretch}{1.5}
\begin{align*}
\begin{array}{c|c}
\multicolumn{2}{c}{\bm{\psi^2 X H^{3} + \textbf{h.c.}}} \\
\hline
 \op{8}{\ell e W H^3}{(1)} & (\bar \ell_{p} \sigma^{\mu\nu} \tau^{a} e_{r} H) (H^\dag H) W^{a}_{\mu\nu} \\
 \op{8}{\ell e W H^3}{(2)} & (\bar \ell_{p} \sigma^{\mu\nu} e_{r} H) (H^\dag \tau^a H) W^{a}_{\mu\nu} \\
 \op{8}{\ell e B H^3}{} & (\bar \ell_{p} \sigma^{\mu\nu}  e_{r} H) (H^\dag H)B_{\mu\nu} \\
 \op{8}{q u G H^3}{} & (\bar q_{p} \sigma^{\mu\nu} T^{\mathscr{A}} u_{r} \tilde H) (H^\dag H) G^{\mathscr{A}}_{\mu\nu} \\
 \op{8}{q u W H^3}{(1)} & (\bar q_{p} \sigma^{\mu\nu} \tau^{a} u_{r} \tilde H) (H^\dag H) W^{a}_{\mu\nu} \\
 \op{8}{q u W H^3}{(2)} & (\bar q_{p} \sigma^{\mu\nu} u_{r} \tilde H) (H^\dag \tau^{a} H) W^{a}_{\mu\nu} \\
 \op{8}{q u B H^3}{} & (\bar q_{p} \sigma^{\mu\nu} u_{r} \tilde H) (H^\dag H) B_{\mu\nu} \\
 \op{8}{q d G H^3}{} & (\bar q_{p} \sigma^{\mu\nu} T^{\mathscr{A}} d_{r} H) (H^\dag H) G^{\mathscr{A}}_{\mu\nu} \\
 \op{8}{q d W H^3}{(1)} & (\bar q_{p} \sigma^{\mu\nu} \tau^{a} d_{r} H) (H^\dag H) W^{a}_{\mu\nu} \\
 \op{8}{q d W H^3}{(2)} & (\bar q_{p} \sigma^{\mu\nu} d_{r} H) (H^\dag \tau^{a} H) W^{a}_{\mu\nu} \\
 \op{8}{q d B H^3}{} & (\bar q_{p} \sigma^{\mu\nu} d_{r} H) (H^\dag H) B_{\mu\nu} \\
\end{array}
\end{align*}
\end{minipage}
\hspace{1cm}
\begin{minipage}[t]{2.5cm}
\vspace{-0.5cm}
\renewcommand{\arraystretch}{1.5}
\renewcommand{\arraystretch}{1.5}
\begin{align*}
\begin{array}{c|c}
\multicolumn{2}{c}{\bm{\psi^2 H^4 D} } \\
\hline
 \op{8}{\ell^2H^4 D}{(1)}   & (\bar \ell_{p} \gamma^{\mu} \ell_{r})\,(H^\dag i \overset{\leftrightarrow}{D}_{\mu} H) (H^\dag H) \\
 \op{8}{\ell^2H^4 D}{(2)}   & (\bar \ell_{p} \gamma^{\mu} \tau^{a} \ell_{r})\,(H^\dag i \overset{\leftrightarrow}{D}_{\mu} H) (H^\dag \tau^{a} H)  \\
 \op{8}{\ell^2H^4 D}{(3)}   & (\bar \ell_{p} \gamma^{\mu} \tau^{a} \ell_{r})\,(H^\dag i \overset{\leftrightarrow}{D}_{\mu}^{a} H) (H^\dag H)  \\
 \op{8}{\ell^2H^4 D}{(4)}   & \epsilon_{abc}(\bar \ell_{p} \gamma^{\mu} \tau^{a} \ell_{r})\,(H^\dag i \overset{\leftrightarrow}{D}_{\mu}^{b} H) (H^\dag \tau^{c} H)  \\
 \op{8}{e^2H^4 D}{}   & (\bar e_{p} \gamma^{\mu} e_{r})\,(H^\dag i \overset{\leftrightarrow}{D}_{\mu} H)  (H^\dag H) \\
 \op{8}{q^2H^4 D}{(1)}   & (\bar q_{p} \gamma^{\mu} q_{r})\,(H^\dag i \overset{\leftrightarrow}{D}_{\mu} H)  (H^\dag H) \\
 \op{8}{q^2H^4 D}{(2)}   & (\bar q_{p} \gamma^{\mu} \tau^{a} q_{r})\,(H^\dag i \overset{\leftrightarrow}{D}_{\mu} H) (H^\dag \tau^{a} H)  \\
 \op{8}{q^2H^4 D}{(3)}   & (\bar q_{p} \gamma^{\mu} \tau^{a} q_{r})\,(H^\dag i \overset{\leftrightarrow}{D}_{\mu}^{a} H) (H^\dag H)  \\
 \op{8}{q^2H^4 D}{(4)}   & \epsilon_{abc}(\bar q_{p} \gamma^{\mu} \tau^{a} q_{r})\,(H^\dag i \overset{\leftrightarrow}{D}_{\mu}^{b} H) (H^\dag \tau^{c} H)  \\
 \op{8}{u^2H^4 D}{}   & (\bar u_{p} \gamma^{\mu} u_{r})\,(H^\dag i \overset{\leftrightarrow}{D}_{\mu} H) (H^\dag H)  \\
 \op{8}{d^2H^4 D}{}   & (\bar d_{p} \gamma^{\mu} d_{r})\,(H^\dag i \overset{\leftrightarrow}{D}_{\mu} H) (H^\dag H)  \\
 \op{8}{udH^4 D}{}   & (\bar u_{p} \gamma^{\mu} d_{r})\, (\tilde H^\dag i D_{\mu} H) (H^\dag H) + \textrm{h.c.}  \\
\end{array}
\end{align*}
\end{minipage}
\end{center}
\caption{\label{dim:8_fermion} Fermionic dimension-eight operators in the SMEFT (not including four-fermion operators or other operators not appearing in the initial Lagrangian). This notation coincides for the most part with the notation used in ref.~\cite{Murphy:2020rsh}.}
\end{table}
%

\section{Renormalization Group Evolution in the SMEFT to Dimension Eight}\label{sec:RGEresults}

We now list the renormalization group equations for bosonic operators renormalized by fermionic operators in the SMEFT up to dimension eight. There is partial overlap with the results in refs.~\cite{Chala:2021pll,DasBakshi:2022mwk,AccettulliHuber:2021uoa}.

\subsection{Definitions}

Here we list some combinations of couplings that enter in the RGE results below.
\begin{align}
	\gamma_{H}^{(Y)} =& 
	{\rm Tr}\left[ Y^{\dagger}_{e} Y_{e} +  N_{c} Y^{\dagger}_{u} Y_{u} + N_{c} Y^{\dagger}_{d} Y_{d} \right] \,.
\end{align}
For the terms linear in $\coef{6}{\psi^2 H^2 D}{}$ or  $\coef{8}{\psi^2 H^4 D}{}$, we define 
\begin{align}
	\kappaHypKinSix =& \left[ y_{e} \; \coef{6}{\underset{tt}{e^2 H^2 D}}{} + 2 y_{\ell} \; \coef{6}{\underset{tt}{\ell^2 H^2 D}}{(1)}  + N_{c} y_{u}  \; \coef{6}{\underset{tt}{u^2 H^2 D}}{} + N_{c} y_{d}  \; \coef{6}{\underset{tt}{d^2 H^2 D}}{}  + 2 N_{c} y_{q}  \;  \coef{6}{\underset{tt}{q^2 H^2 D}}{(1)} \right] \,,  \\
    \kappaHypKinEight =& \left[ y_{e} \; \coef{8}{\underset{tt}{e^2 H^4 D}}{} + 2 y_{\ell} \; \coef{8}{\underset{tt}{\ell^2 H^4 D}}{(1)} + N_{c} y_{u}  \; \coef{8}{\underset{tt}{u^2 H^4 D}}{} + N_{c} y_{d}  \; \coef{8}{\underset{tt}{d^2 H^4 D}}{}  + 2 N_{c} y_{q}  \; \coef{8}{\underset{tt}{q^2 H^4 D}}{(1)}  \right] \,,  \\
    \kappaPauliKinSix =& \left[ \coef{6}{\underset{tt}{\ell^2 H^2 D}}{(3)}  + N_{c} \; \coef{6}{\underset{tt}{q^2 H^2 D}}{(3)} \right] \,, \\
	\kappaPauliKinEight =& \left[ \coef{8}{\underset{tt}{\ell^2 H^4 D}}{(3)}  + N_{c} \; \coef{8}{\underset{tt}{q^2 H^4 D}}{(3)}  \right] \,, \\
    \kappaPauliKinEightTwo =& \left[ \coef{8}{\underset{tt}{\ell^2 H^4 D}}{(2)}  + N_{c} \; \coef{8}{\underset{tt}{q^2 H^4 D}}{(2)}  \right] \,,
\end{align}
and for the terms quadratic in $\coef{6}{\psi^2 H^2 D}{}$ we use 
\begin{align}
	\kappaKinSixKinSix =& {\rm Tr}\left[ \left( \coef{6}{e^2 H^2 D}{} \right)^{2} + 2 \left(\coef{6}{\ell^2 H^2 D}{(1)}  \right)^{2} + N_{c} \left( \coef{6}{u^2 H^2 D}{} \right)^{2} + N_{c} \left( \coef{6}{d^2 H^2 D}{} \right)^{2} 
    \right. \nonumber \\ & \left. \qquad
    + 2 N_{c} \left( \coef{6}{q^2 H^2 D}{(1)} \right)^{2} \right] \,, \\
	\kappaKinSixKinSixSUTwo =& {\rm Tr}\left[ 2 \left( \coef{6}{\ell^2 H^2 D}{(3)} \right)^{2}  + 2 N_{c} \left( \coef{6}{q^2 H^2 D}{(3)} \right)^{2} \right] \,, \\
	\kappaKinSixKinSixud =& {\rm Tr}\left[ 2 N_{c} \left(\coef{6}{udH^2 D}{}  \right) \left( \coef{6}{udH^2 D}{\dagger} \right) \right] \,.
\end{align}
For the terms involving the Yukawa couplings, we define 
\begin{align}
    \kappaYukawaFourYukawaSix =& {\rm Tr}\left[ Y_{e} \,\,\coef{6}{\ell e H^3}{} + N_c Y_{u} \,\,\coef{6}{quH^3}{} + N_c Y_{d} \,\,\coef{6}{qdH^3}{}  + {\rm h.c.} \right] \,, \\ 
    \kappaYukawaFourYukawaEight =& {\rm Tr}\left[ Y_{e} \,\,\coef{8}{\ell e H^5}{} + N_c Y_{u} \,\,\coef{8}{quH^5}{} + N_c Y_{d} \,\,\coef{8}{qdH^5}{}  + {\rm h.c.} \right] \,, \\ 
    \kappaYukawaSixYukawaSix =& {\rm Tr}\left[ \coef{6}{\ell e H^3}{} \,\,\coef{6}{\ell e H^3}{\dagger} + N_c \,\,\coef{6}{quH^3}{} \,\,\coef{6}{quH^3}{\dagger} + N_c \,\,\coef{6}{qdH^3}{} \,\,\coef{6}{qdH^3}{\dagger}  \right] \,, \\ 
	\kappaYukawaKinSix =& {\rm Tr}\left[- Y_{e} Y^{\dagger}_{e} \coef{6}{e^2 H^2 D}{} + Y^{\dagger}_{e} Y_{e} \coef{6}{\ell^2 H^2 D}{(1)} - N_c Y_{d} Y^{\dagger}_{d} \coef{6}{d^2 H^2 D}{} + N_c Y^{\dagger}_{d} Y_{d} \coef{6}{q^2 H^2 D}{(1)}  \right.\nonumber \\ & \left. \qquad + N_c  Y_{u} Y^{\dagger}_{u}  \coef{6}{u^2 H^2 D}{} - N_c Y^{\dagger}_{u} Y_{u}  \coef{6}{q^2 H^2 D}{(1)} \right] \,, \\
 	\kappaYukawaKinEight =& {\rm Tr}\left[- Y_{e} Y^{\dagger}_{e} \coef{8}{e^2 H^4 D}{} + Y^{\dagger}_{e} Y_{e} \coef{8}{\ell^2 H^4 D}{(1)} - N_c Y_{d} Y^{\dagger}_{d} \coef{8}{d^2 H^4 D}{} + N_c Y^{\dagger}_{d} Y_{d} \coef{8}{q^2 H^4 D}{(1)}  \right.\nonumber \\ & \left. \qquad + N_c  Y_{u} Y^{\dagger}_{u}  \coef{8}{u^2 H^4 D}{} - N_c Y^{\dagger}_{u} Y_{u}  \coef{8}{q^2 H^4 D}{(1)} \right] \,, \\
 	\kappaYukawaKinThreeSix =& {\rm Tr}\left[ Y^{\dagger}_{e} Y_{e}  \coef{6}{\ell^2 H^2 D}{(3)} + N_c \left( Y^{\dagger}_{d} Y_{d} + Y^{\dagger}_{u} Y_{u} \right) \coef{6}{q^2 H^2 D}{(3)} \right] \,, \\
   	\kappaYukawaKinThreeEight =& {\rm Tr}\left[ Y^{\dagger}_{e} Y_{e}  \coef{8}{\ell^2 H^4 D}{(3)} + N_c \left( Y^{\dagger}_{d} Y_{d} + Y^{\dagger}_{u} Y_{u} \right) \coef{8}{q^2 H^4 D}{(3)} \right] \,, \\
    \kappaYukawaKinudSix =& {\rm Tr}\left[ - N_c Y_{d} Y^{\dagger}_{u} \coef{6}{udH^2 D}{}  - N_c Y_{u} Y^{\dagger}_{d} \coef{6}{udH^2 D}{\dagger} \right] \,, \\
    \kappaYukawaKinudEight =& {\rm Tr}\left[ - N_c Y_{d} Y^{\dagger}_{u} \coef{8}{udH^4 D}{}  - N_c Y_{u} Y^{\dagger}_{d} \coef{8}{udH^4 D}{\dagger} \right] \,, \\
    \kappaYukawaKinTwoEight =& {\rm Tr}\left[ Y^{\dagger}_{e} Y_{e}  \coef{8}{\ell^2 H^4 D}{(2)} + N_c \left( Y^{\dagger}_{d} Y_{d} + Y^{\dagger}_{u} Y_{u} \right) \coef{8}{q^2 H^4 D}{(2)} \right] \,, \\
	\kappaYukawaKinSixKinSix =& {\rm Tr}\left[ 2 Y_{e} \coef{6}{\ell^2 H^2 D}{(1)}  Y^{\dagger}_{e} \coef{6}{e^2 H^2 D}{} - Y_{e} Y^{\dagger}_{e} \coef{6}{e^2 H^2 D}{}\coef{6}{e^2 H^2 D}{} - Y^{\dagger}_{e} Y_{e} \coef{6}{\ell^2 H^2 D}{(1)}\coef{6}{\ell^2 H^2 D}{(1)} \right. \nonumber \\ & \left. \quad + 2 N_c Y_{d} \coef{6}{q^2 H^2 D}{(1)}  Y^{\dagger}_{d} \coef{6}{d^2 H^2 D}{} - N_c Y_{d} Y^{\dagger}_{d} \coef{6}{d^2 H^2 D}{}\coef{6}{d^2 H^2 D}{} - N_c Y^{\dagger}_{d} Y_{d} \coef{6}{q^2 H^2 D}{(1)}\coef{6}{q^2 H^2 D}{(1)} \right. \nonumber \\ & \left.  \quad + 2 N_c Y_{u} \coef{6}{q^2 H^2 D}{(1)}  Y^{\dagger}_{u} \coef{6}{u^2 H^2 D}{} - N_c Y_{u} Y^{\dagger}_{u} \coef{6}{u^2 H^2 D}{}\coef{6}{u^2 H^2 D}{} - N_c Y^{\dagger}_{u} Y_{u} \coef{6}{q^2 H^2 D}{(1)}\coef{6}{q^2 H^2 D}{(1)} \right] \,, \\
    \kappaYukawaKinSixKinSixThree =& {\rm Tr}\left[ 2 Y_{e} \coef{6}{\ell^2 H^2 D}{(3)}  Y^{\dagger}_{e} \coef{6}{e^2 H^2 D}{}  - Y^{\dagger}_{e} Y_{e} \left( \coef{6}{\ell^2 H^2 D}{(1)}\coef{6}{\ell^2 H^2 D}{(3)} + \coef{6}{\ell^2 H^2 D}{(3)}\coef{6}{\ell^2 H^2 D}{(1)}  \right) \right. \nonumber \\ & \left. \quad + 2 N_c Y_{d} \coef{6}{q^2 H^2 D}{(3)}  Y^{\dagger}_{d} \coef{6}{d^2 H^2 D}{}  - N_c Y^{\dagger}_{d} Y_{d} \left( \coef{6}{q^2 H^2 D}{(1)}\coef{6}{q^2 H^2 D}{(3)} + \coef{6}{q^2 H^2 D}{(3)}\coef{6}{q^2 H^2 D}{(1)}  \right) \right. \nonumber \\ & \left.  \quad - 2 N_c Y_{u} \coef{6}{q^2 H^2 D}{(3)}  Y^{\dagger}_{u} \coef{6}{u^2 H^2 D}{} + N_c Y^{\dagger}_{u} Y_{u} \left( \coef{6}{q^2 H^2 D}{(1)}\coef{6}{q^2 H^2 D}{(3)} +  \coef{6}{q^2 H^2 D}{(3)}\coef{6}{q^2 H^2 D}{(1)}  \right) \right] \,, \\ 
    \kappaYukawaKinSixThreeKinSixThree =& {\rm Tr}\left[   - Y^{\dagger}_{e} Y_{e} \coef{6}{\ell^2 H^2 D}{(3)}\coef{6}{\ell^2 H^2 D}{(3)}   - N_c Y^{\dagger}_{d} Y_{d} \coef{6}{q^2 H^2 D}{(3)}\coef{6}{q^2 H^2 D}{(3)} - N_c Y^{\dagger}_{u} Y_{u}  \coef{6}{q^2 H^2 D}{(3)}\coef{6}{q^2 H^2 D}{(3)}  \right] \,, \\ 
    \kappaYukawaSixKinSix =& {\rm Tr}\left[- \left( Y_{e} \coef{6}{\ell e H^3}{}  + \coef{6}{\ell e H^3}{\dagger} Y^{\dagger}_{e} \right) \coef{6}{e^2 H^2 D}{} + \left( Y^{\dagger}_{e} \coef{6}{\ell e H^3}{\dagger}  + \coef{6}{\ell e H^3}{} Y_{e} \right) \coef{6}{\ell^2 H^2 D}{(1)} \right. \nonumber \\ & \left. \quad - N_c \left( Y_{d} \coef{6}{q d H^3}{}  + \coef{6}{q d H^3}{\dagger} Y^{\dagger}_{d} \right) \coef{6}{d^2 H^2 D}{} + N_c \left( Y^{\dagger}_{d} \coef{6}{q d H^3}{\dagger}  + \coef{6}{q d H^3}{} Y_{d} \right) \coef{6}{q^2 H^2 D}{(1)} \right. \nonumber \\ & \left. \quad + N_c \left( Y_{u} \coef{6}{q u H^3}{}  + \coef{6}{q u H^3}{\dagger} Y^{\dagger}_{u} \right) \coef{6}{u^2 H^2 D}{} - N_c \left( Y^{\dagger}_{u} \coef{6}{q u H^3}{\dagger}  + \coef{6}{q u H^3}{} Y_{u} \right) \coef{6}{q^2 H^2 D}{(1)} \right] \,, \\
    \kappaYukawaSixKinSixThree =& {\rm Tr}\left[ \left( Y^{\dagger}_{e} \coef{6}{\ell e H^3}{\dagger}  + \coef{6}{\ell e H^3}{} Y_{e} \right) \coef{6}{\ell^2 H^2 D}{(3)} \right. \nonumber \\ & \left. \quad + N_c \left( Y^{\dagger}_{d} \coef{6}{q d H^3}{\dagger}  + \coef{6}{q d H^3}{} Y_{d} \right) \coef{6}{q^2 H^2 D}{(3)}   + N_c \left( Y^{\dagger}_{u} \coef{6}{q u H^3}{\dagger}  + \coef{6}{q u H^3}{} Y_{u} \right) \coef{6}{q^2 H^2 D}{(3)} \right] \,, \\
    \kappaYukawaSixKinSixud =& {\rm Tr}\left[ N_c \left( Y_{d} \coef{6}{q u H^3}{}  + \coef{6}{q d H^3}{\dagger} Y^{\dagger}_{u} \right) \coef{6}{ud H^2 D}{} + {\rm h.c. }  \right] \,, \\
    \kappaYukawaKinSixudKinSixud =& {\rm Tr}\left[ N_c \left( Y_{u} Y^{\dagger}_{u} \right) \coef{6}{ud H^2 D}{} \coef{6}{ud H^2 D}{\dagger} + N_c \left( Y_{d} Y^{\dagger}_{d} \right) \coef{6}{ud H^2 D}{\dagger} \coef{6}{ud H^2 D}{}  \right] \,, \\
    \kappaYukawaKinSixThreeKinSixud =& {\rm Tr}\left[ N_c \left( Y_{d} \coef{6}{q^2 H^2 D}{(3)} Y^{\dagger}_{u} \right) \coef{6}{ud H^2 D}{}  + N_c \left( Y_{u} \coef{6}{q^2 H^2 D}{(3)} Y^{\dagger}_{d} \right) \coef{6}{ud H^2 D}{\dagger} \right] \,, \\
    \kappaYukawaYYYyukawaSix =& {\rm Tr}\left[ Y_e Y^\dagger_e Y_e \;\coef{6}{\ell e H^3}{} + N_{c} Y_u Y^\dagger_u Y_u \;\coef{6}{qu H^3}{} + N_{c} Y_d Y^\dagger_d Y_d \;\coef{6}{qd H^3}{} + {\rm  h.c.} \right] \,, \\
    \kappaYukawaYYYyukawaEight =& {\rm Tr}\left[ Y_e Y^\dagger_e Y_e \;\coef{8}{\ell e H^5}{} + N_{c} Y_u Y^\dagger_u Y_u \;\coef{8}{qu H^5}{} + N_{c} Y_d Y^\dagger_d Y_d \;\coef{8}{qd H^5}{} + {\rm  h.c.} \right] \,, \\
    \kappaYukawaYYyukawaSixYukawaSix =& {\rm Tr}\left[ Y_e \;\coef{6}{\ell e H^3}{} Y_{e} \;\coef{6}{\ell e H^3}{} + N_{c} Y_u \;\coef{6}{qu H^3}{} Y_{u} \;\coef{6}{qu H^3}{} + N_{c} Y_d \;\coef{6}{qd H^3}{} Y_{d} \;\coef{6}{qd H^3}{} 
    \right. \nonumber \\ & \left. \quad
    +2 Y_e \;\coef{6}{\ell e H^3}{} \;\coef{6}{\ell e H^3}{\dagger} Y^{\dagger}_{e} + 2 N_{c} Y_u \;\coef{6}{qu H^3}{} \;\coef{6}{qu H^3}{\dagger} Y^{\dagger}_{u} + 2 Y_d \;\coef{6}{qd H^3}{} \;\coef{6}{qd H^3}{\dagger} Y^{\dagger}_{d} + {\rm h.c.}
    \right] \,.
\end{align}
For the dipole terms we use
\begin{align}
    \tauGG =& \frac{1}{2}{\rm Tr} \left[ \coef{6}{q d G H}{} \coef{6}{q d  GH}{\dagger} + \coef{6}{q u G H}{} \coef{6}{q u  GH}{\dagger} \right] \,, \\
    \tauBB =& {\rm Tr} \left[ \coef{6}{\ell e B H}{} \coef{6}{\ell e BH}{\dagger} + N_c \coef{6}{q d B H}{} \coef{6}{q d  BH}{\dagger} + N_c \coef{6}{q u B H}{} \coef{6}{q u  BH}{\dagger} \right] \,, \\
    \tauBW =& \frac{1}{2}{\rm Tr} \left[ \coef{6}{\ell e B H}{} \coef{6}{\ell e WH}{\dagger} + N_c \coef{6}{q d B H}{} \coef{6}{q d WH}{\dagger} - N_c \coef{6}{q u B H}{} \coef{6}{q u WH}{\dagger}  + {\rm h.c. }\right] \,, \\
    \tauBWp =& \frac{1}{2}{\rm Tr} \left[i \coef{6}{\ell e B H}{} \coef{6}{\ell e WH}{\dagger} +i N_c \coef{6}{q d B H}{} \coef{6}{q d WH}{\dagger} +i N_c \coef{6}{q u B H}{} \coef{6}{q u WH}{\dagger}  + {\rm h.c. }\right] \,, \\
    \tauBWpp =& {\rm Tr} \left[i \coef{6}{\ell e B H}{} \coef{6}{\ell e WH}{\dagger} +i N_c \coef{6}{q d B H}{} \coef{6}{q d WH}{\dagger} -i N_c \coef{6}{q u B H}{} \coef{6}{q u WH}{\dagger}  + {\rm h.c. }\right] \,, \\
    \tauWW =& {\rm Tr} \left[ \coef{6}{\ell e W H}{} \coef{6}{\ell e WH}{\dagger} + N_c \coef{6}{q d W H}{} \coef{6}{q d WH}{\dagger} + N_c \coef{6}{q u W H}{} \coef{6}{q u WH}{\dagger} \right] \,, \\
    \tauWWp =& {\rm Tr} \left[ \coef{6}{\ell e W H}{} \coef{6}{\ell e WH}{\dagger} + N_c \coef{6}{q d W H}{} \coef{6}{q d WH}{\dagger} - N_c \coef{6}{q u W H}{} \coef{6}{q u WH}{\dagger} \right] \,, \\
    \tauYukDipoleBBSix =& {\rm Tr}\left[ (y_e + y_{\ell}) Y_{e} \coef{6}{\ell e B H}{} +  (y_d + y_{q}) N_c Y_{d} \coef{6}{q d B H}{} +  (y_u + y_{q}) N_c Y_{u} \coef{6}{q u B H}{} + {\rm h.c.} \right] \,, \\
    \tauYukDipoleBBEight =& {\rm Tr}\left[ (y_e + y_{\ell}) (Y_{e} \coef{8}{\ell e B H^3}{} - \coef{6}{\ell e H^3}{\dagger} \coef{6}{\ell e B H}{} ) +  (y_d + y_{q}) N_c ( Y_{d} \coef{8}{q d B H^3}{} - \coef{6}{qd H^3}{\dagger} \coef{6}{q d B H}{}) \right.\nonumber\\ &\left. \qquad +  (y_u + y_{q}) N_c (Y_{u} \coef{8}{q u B H^3}{} - \coef{6}{q u H^3}{\dagger} \coef{6}{q u B H}{}) + {\rm h.c.} \right] \,, \\
    \tauYukDipoleBBtildeSix =& {\rm Tr}\left[ i(y_e + y_{\ell}) Y_{e} \coef{6}{\ell e B H}{} +  i(y_d + y_{q}) N_c Y_{d} \coef{6}{q d B H}{} +  i(y_u + y_{q}) N_c Y_{u} \coef{6}{q u B H}{} + {\rm h.c.} \right] \,, \\
    \tauYukDipoleBBtildeEight =& {\rm Tr}\left[ i(y_e + y_{\ell}) (Y_{e} \coef{8}{\ell e B H^3}{} - \coef{6}{\ell e H^3}{\dagger} \coef{6}{\ell e B H}{} ) +  i(y_d + y_{q}) N_c ( Y_{d} \coef{8}{q d B H^3}{} - \coef{6}{qd H^3}{\dagger} \coef{6}{q d B H}{}) \right.\nonumber\\ &\left. \qquad +  i(y_u + y_{q}) N_c (Y_{u} \coef{8}{q u B H^3}{} - \coef{6}{q u H^3}{\dagger} \coef{6}{q u B H}{}) + {\rm h.c.} \right] \,, \\
    \tauYukDipoleBWSix =& {\rm Tr}\left[ (y_e + y_{\ell}) Y_{e} \coef{6}{\ell e W H}{} +  (y_d + y_{q}) N_c Y_{d} \coef{6}{q d W H}{} -  (y_u + y_{q}) N_c Y_{u} \coef{6}{q u W H}{} + {\rm h.c.} \right] \,, \\
    \tauYukDipoleBWEight =& {\rm Tr}\left[ (y_e + y_{\ell}) (Y_{e} \coef{8}{\ell e W H^3}{(1)} + Y_{e} \coef{8}{\ell e W H^3}{(2)} - \coef{6}{\ell e H^3}{\dagger} \coef{6}{\ell e W H}{}) \right.\nonumber \\ & \left. \qquad +  (y_d + y_{q}) N_c (Y_{d} \coef{8}{q d W H^3}{(1)} + Y_{d} \coef{8}{q d W H^3}{(2)} - \coef{6}{q d H^3}{\dagger} \coef{6}{q d W H}{}) \right. \nonumber \\ & \left. \qquad -  (y_u + y_{q}) N_c (Y_{u} \coef{8}{q u W H^3}{(1)} - Y_{u} \coef{8}{q u W H^3}{(2)} - \coef{6}{q u H^3}{\dagger} \coef{6}{q u W H}{}) + {\rm h.c.} \right] \,, \\
    \tauYukDipoleBWtildeSix =& {\rm Tr}\left[ i(y_e + y_{\ell}) Y_{e} \coef{6}{\ell e W H}{} +  i(y_d + y_{q}) N_c Y_{d} \coef{6}{q d W H}{} -  i(y_u + y_{q}) N_c Y_{u} \coef{6}{q u W H}{} + {\rm h.c.} \right] \,, \\
    \tauYukDipoleBWtildeEight =& {\rm Tr}\left[ i(y_e + y_{\ell}) (Y_{e} \coef{8}{\ell e W H^3}{(1)} + Y_{e} \coef{8}{\ell e W H^3}{(2)} - \coef{6}{\ell e H^3}{\dagger} \coef{6}{\ell e W H}{}) \right.\nonumber \\ & \left. \qquad +  i(y_d + y_{q}) N_c (Y_{d} \coef{8}{q d W H^3}{(1)} + Y_{d} \coef{8}{q d W H^3}{(2)} - \coef{6}{q d H^3}{\dagger} \coef{6}{q d W H}{}) \right. \nonumber \\ & \left. \qquad -  i(y_u + y_{q}) N_c (Y_{u} \coef{8}{q u W H^3}{(1)} - Y_{u} \coef{8}{q u W H^3}{(2)} - \coef{6}{q u H^3}{\dagger} \coef{6}{q u W H}{}) + {\rm h.c.} \right] \,, \\
    \tauYukDipoleWBSix =& {\rm Tr} \left[ Y_e \coef{6}{\ell e B H}{} + N_c Y_d \coef{6}{q d B H}{} - N_c Y_u \coef{6}{q u B H}{} + {\rm h.c.}\right] \,, \\
    \tauYukDipoleWBEight =& {\rm Tr} \left[ (Y_e \coef{8}{\ell e B H^3}{} - \coef{6}{\ell e H^3}{\dagger} \coef{6}{\ell e B H}{}) + N_c (Y_d \coef{8}{q d B H^3}{} - \coef{6}{q d H^3}{\dagger} \coef{6}{q d B H}{})  \right. \nonumber \\ & \left. \qquad - N_c (Y_u \coef{8}{q u B H^3}{} - \coef{6}{q u H^3}{} \coef{6}{q u B H}{}) + {\rm h.c.}\right] \,, \\
    \tauYukDipoleWtildeBSix =& {\rm Tr} \left[i Y_e \coef{6}{\ell e B H}{} + i N_c Y_d \coef{6}{q d B H}{} - i N_c Y_u \coef{6}{q u B H}{} + {\rm h.c.}\right] \,, \\
    \tauYukDipoleWtildeBEight =& {\rm Tr} \left[ i(Y_e \coef{8}{\ell e B H^3}{} - \coef{6}{\ell e H^3}{\dagger} \coef{6}{\ell e B H}{}) + iN_c (Y_d \coef{8}{q d B H^3}{} - \coef{6}{q d H^3}{\dagger} \coef{6}{q d B H}{})  \right. \nonumber \\ & \left. \qquad - iN_c (Y_u \coef{8}{q u B H^3}{} - \coef{6}{q u H^3}{\dagger} \coef{6}{q u B H}{}) + {\rm h.c.}\right] \,, \\
    \tauYukDipoleWWSix =& {\rm Tr} \left[ Y_e \coef{6}{\ell e W H}{} + N_c Y_d \coef{6}{q d W H}{} + N_c Y_u \coef{6}{q u W H}{} + {\rm h.c.}\right] \,, \\
    \tauYukDipoleWWEight =& {\rm Tr} \left[ (Y_e \coef{8}{\ell e W H^3}{(1)} - \coef{6}{\ell e H^3}{\dagger} \coef{6}{\ell e W H}{}) + N_c (Y_d \coef{8}{q d W H^3}{(1)} - \coef{6}{q d H^3}{\dagger} \coef{6}{q d W H}{}) \right. \nonumber \\ & \left. \qquad + N_c (Y_u \coef{8}{q u W H^3}{(1)} - \coef{6}{q u H^3}{\dagger} \coef{6}{q u W H}{}) + {\rm h.c.}\right] \,, \\
    \tauYukDipoleWtildeWSix =& {\rm Tr} \left[ i Y_e \coef{6}{\ell e W H}{} + i N_c Y_d \coef{6}{q d W H}{} + i N_c Y_u \coef{6}{q u W H}{} + {\rm h.c.}\right] \,, \\
    \tauYukDipoleWtildeWEight =& {\rm Tr} \left[ i(Y_e \coef{8}{\ell e W H^3}{(1)} - \coef{6}{\ell e H^3}{\dagger} \coef{6}{\ell e W H}{}) + iN_c (Y_d \coef{8}{q d W H^3}{(1)} - \coef{6}{q d H^3}{\dagger} \coef{6}{q d W H}{}) \right. \nonumber \\ & \left. \qquad + iN_c (Y_u \coef{8}{q u W H^3}{(1)} - \coef{6}{q u H^3}{\dagger} \coef{6}{q u W H}{}) + {\rm h.c.}\right] \,, \\
    \tauYukDipoleWWtwoEight =& {\rm Tr} \left[ Y_e \coef{8}{\ell e W H^3}{(2)} + N_c Y_d \coef{8}{q d W H^2}{(2)} - N_c Y_u \coef{8}{q u W H^3}{(2)} + {\rm h.c.}\right] \,, \\
    \tauYukDipoleWtildeWtwoEight =& {\rm Tr} \left[ iY_e \coef{8}{\ell e W H^3}{(2)} + iN_c Y_d \coef{8}{q d W H^2}{(2)} - iN_c Y_u \coef{8}{q u W H^3}{(2)} + {\rm h.c.}\right] \,, \\
    \tauYukDipoleGGSix =& {\rm Tr} \left[ Y_d \coef{6}{q d G H}{}  + Y_u \coef{6}{q u G H}{}  + {\rm h.c.} \right] \,, \\
    \tauYukDipoleGGEight =& {\rm Tr} \left[ Y_d \coef{8}{q d G H^3}{} - \coef{6}{q d H^3}{\dagger} \coef{6}{q d G H}{}  + Y_u \coef{8}{q u G H^3}{} - \coef{6}{q u H^3}{\dagger} \coef{6}{q u G H}{}  + {\rm h.c.} \right] \,, \\
    \tauYukDipoleGGtildeSix =& {\rm Tr} \left[ i Y_d \coef{6}{q d G H}{}  + i Y_u \coef{6}{q u G H}{}  + {\rm h.c.} \right] \,, \\
    \tauYukDipoleGGtildeEight =& {\rm Tr} \left[ i Y_d \coef{8}{q d G H^3}{} - i \coef{6}{q d H^3}{\dagger} \coef{6}{q d G H}{}  + i Y_u \coef{8}{q u G H^3}{} - i \coef{6}{q u H^3}{\dagger} \coef{6}{q u G H}{}  + {\rm h.c.} \right] \,. 
\end{align}
We also need
\begin{align}
    \tauYukKinSixDipoleB =& {\rm Tr} \left[ \coef{6}{e^2 H^2 D}{(1)} Y_e \coef{6}{\ell e B H}{} + \coef{6}{\ell^2 H^2 D}{(1)} \coef{6}{\ell e B H}{} Y_e + N_c \coef{6}{d^2 H^2 D}{(1)} Y_d \coef{6}{qd B H}{} \right.\nonumber \\ & \left. \qquad + N_c \coef{6}{u^2 H^2 D}{(1)} Y_u \coef{6}{qu B H}{}  + N_c \coef{6}{q^2 H^2 D}{(1)} (\coef{6}{qd B H}{} Y_d +  \coef{6}{qu B H}{} Y_u ) + {\rm h.c.} \right] \,, \\
    \tauYukKinSixDipoleBtilde =& {\rm Tr} \left[ i\coef{6}{e^2 H^2 D}{(1)} Y_e \coef{6}{\ell e B H}{} + i\coef{6}{\ell^2 H^2 D}{(1)} \coef{6}{\ell e B H}{} Y_e + iN_c \coef{6}{d^2 H^2 D}{(1)} Y_d \coef{6}{qd B H}{} \right.\nonumber \\ & \left. \qquad + iN_c \coef{6}{u^2 H^2 D}{(1)} Y_u \coef{6}{qu B H}{}  + iN_c \coef{6}{q^2 H^2 D}{(1)} (\coef{6}{qd B H}{} Y_d +  \coef{6}{qu B H}{} Y_u ) + {\rm h.c.} \right] \,, \\
    \tauYukKinSixDipoleW =& {\rm Tr} \left[ \coef{6}{e^2 H^2 D}{(1)} Y_e \coef{6}{\ell e W H}{} + \coef{6}{\ell^2 H^2 D}{(1)} \coef{6}{\ell e W H}{} Y_e + N_c \coef{6}{d^2 H^2 D}{(1)} Y_d \coef{6}{qd W H}{} \right.\nonumber \\ & \left. \qquad - N_c \coef{6}{u^2 H^2 D}{(1)} Y_u \coef{6}{qu W H}{}  + N_c \coef{6}{q^2 H^2 D}{(1)} (\coef{6}{qd W H}{} Y_d -  \coef{6}{qu W H}{} Y_u ) + {\rm h.c.} \right] \,, \\
    \tauYukKinSixDipoleWtilde =& {\rm Tr} \left[ i\coef{6}{e^2 H^2 D}{(1)} Y_e \coef{6}{\ell e W H}{} + i\coef{6}{\ell^2 H^2 D}{(1)} \coef{6}{\ell e W H}{} Y_e + iN_c \coef{6}{d^2 H^2 D}{(1)} Y_d \coef{6}{qd W H}{} \right.\nonumber \\ & \left. \qquad - iN_c \coef{6}{u^2 H^2 D}{(1)} Y_u \coef{6}{qu W H}{}  + iN_c \coef{6}{q^2 H^2 D}{(1)} (\coef{6}{qd W H}{} Y_d -  \coef{6}{qu W H}{} Y_u ) + {\rm h.c.} \right] \,, \\
    \tauYukKinThreeSixDipoleB =& {\rm Tr} \left[  \coef{6}{\ell^2 H^2 D}{(3)} \coef{6}{\ell e B H}{} Y_e  + N_c \coef{6}{q^2 H^2 D}{(3)} (\coef{6}{qd B H}{} Y_d -  \coef{6}{qu B H}{} Y_u ) + {\rm h.c.} \right] \,, \\
    \tauYukKinThreeSixDipoleBtilde =& {\rm Tr} \left[  i\coef{6}{\ell^2 H^2 D}{(3)} \coef{6}{\ell e B H}{} Y_e  + iN_c \coef{6}{q^2 H^2 D}{(3)} (\coef{6}{qd B H}{} Y_d -  \coef{6}{qu B H}{} Y_u ) + {\rm h.c.} \right] \,, \\
    \tauYukKinThreeSixDipoleW =& {\rm Tr} \left[  \coef{6}{\ell^2 H^2 D}{(3)} \coef{6}{\ell e W H}{} Y_e  + N_c \coef{6}{q^2 H^2 D}{(3)} (\coef{6}{qd W H}{} Y_d +  \coef{6}{qu W H}{} Y_u ) + {\rm h.c.} \right] \,, \\
    \tauYukKinThreeSixDipoleWtilde =& {\rm Tr} \left[  i\coef{6}{\ell^2 H^2 D}{(3)} \coef{6}{\ell e W H}{} Y_e  + iN_c \coef{6}{q^2 H^2 D}{(3)} (\coef{6}{qd W H}{} Y_d +  \coef{6}{qu W H}{} Y_u ) + {\rm h.c.} \right] \,, \\
    \tauYukKinThreeSixDipoleWeps =& {\rm Tr} \left[  i\coef{6}{\ell^2 H^2 D}{(3)} \coef{6}{\ell e W H}{} Y_e  + iN_c \coef{6}{q^2 H^2 D}{(3)} (\coef{6}{qd W H}{} Y_d -  \coef{6}{qu W H}{} Y_u ) + {\rm h.c.} \right] \,, \\
    \tauYukKinThreeSixDipoleWtildeeps =& {\rm Tr} \left[  \coef{6}{\ell^2 H^2 D}{(3)} \coef{6}{\ell e W H}{} Y_e  + N_c \coef{6}{q^2 H^2 D}{(3)} (\coef{6}{qd W H}{} Y_d -  \coef{6}{qu W H}{} Y_u ) + {\rm h.c.} \right] \,, \\
    \tauYukKinudSixDipoleW =& {\rm Tr} \left[  N_c \coef{6}{ud H^2 D}{} ( Y_d \coef{6}{qu W H}{}  +  \coef{6}{qd W H}{\dagger} Y_u^{\dagger} ) + {\rm h.c.} \right] \,, \\
    \tauYukKinudSixDipoleWtilde =& {\rm Tr} \left[ i N_c \coef{6}{ud H^2 D}{} ( Y_d \coef{6}{qu W H}{}  -  \coef{6}{qd W H}{\dagger} Y_u^{\dagger} ) + {\rm h.c.} \right] \,, \\
    \tauYukKinudSixDipoleWeps =& {\rm Tr} \left[ i N_c \coef{6}{ud H^2 D}{} ( Y_d \coef{6}{qu W H}{}  +  \coef{6}{qd W H}{\dagger} Y_u^{\dagger} ) + {\rm h.c.} \right] \,, \\
    \tauYukKinudSixDipoleWtildeeps =& {\rm Tr} \left[  N_c \coef{6}{ud H^2 D}{} ( Y_d \coef{6}{qu W H}{}  -  \coef{6}{qd W H}{\dagger} Y_u^{\dagger} ) + {\rm h.c.} \right] \,,
\end{align}
and
\begin{align}
    \tauYukTwoDipoleBB =& {\rm Tr}\left[ Y_e \coef{6}{\ell e B H}{} Y_e \coef{6}{\ell e B H}{} + N_{c} Y_d \coef{6}{q d B H}{} Y_d \coef{6}{q d B H}{} + N_{c} Y_u \coef{6}{q u B H}{} Y_u \coef{6}{q u B H}{} + {\rm h.c.}  \right] \,, \\
    \tauYukTwoDipoleBBtilde =& {\rm Tr}\left[ iY_e \coef{6}{\ell e B H}{} Y_e \coef{6}{\ell e B H}{} + iN_{c} Y_d \coef{6}{q d B H}{} Y_d \coef{6}{q d B H}{} + iN_{c} Y_u \coef{6}{q u B H}{} Y_u \coef{6}{q u B H}{} + {\rm h.c.}  \right] \,, \\
    \tauYukTwoDipoleBW =& {\rm Tr}\left[ Y_e \coef{6}{\ell e B H}{} Y_e \coef{6}{\ell e W H}{} + N_{c} Y_d \coef{6}{q d B H}{} Y_d \coef{6}{q d W H}{} - N_{c} Y_u \coef{6}{q u B H}{} Y_u \coef{6}{q u W H}{} + {\rm h.c.}  \right] \,, \\
    \tauYukTwoDipoleBWtilde =& {\rm Tr}\left[ iY_e \coef{6}{\ell e B H}{} Y_e \coef{6}{\ell e W H}{} + iN_{c} Y_d \coef{6}{q d B H}{} Y_d \coef{6}{q d W H}{} - iN_{c} Y_u \coef{6}{q u B H}{} Y_u \coef{6}{q u W H}{} + {\rm h.c.}  \right] \,, \\
    \tauYukTwoDipoleWW =& {\rm Tr}\left[ Y_e \coef{6}{\ell e W H}{} Y_e \coef{6}{\ell e W H}{} + N_{c} Y_d \coef{6}{q d W H}{} Y_d \coef{6}{q d W H}{} + N_{c} Y_u \coef{6}{q u W H}{} Y_u \coef{6}{q u W H}{} + {\rm h.c.}  \right] \,, \\
    \tauYukTwoDipoleWWtilde =& {\rm Tr}\left[ iY_e \coef{6}{\ell e W H}{} Y_e \coef{6}{\ell e W H}{} + iN_{c} Y_d \coef{6}{q d W H}{} Y_d \coef{6}{q d W H}{} + iN_{c} Y_u \coef{6}{q u W H}{} Y_u \coef{6}{q u W H}{} + {\rm h.c.}  \right] \,, \\
    \tauYukTwoDipoleWWtwo =& {\rm Tr}\left[ N_c Y_d \coef{6}{q u W H}{} Y_u \coef{6}{q d G H}{} + {\rm h.c.}  \right] \,, \\
    \tauYukTwoDipoleWWtildetwo =& {\rm Tr}\left[ iN_c Y_d \coef{6}{q u W H}{} Y_u \coef{6}{q d G H}{} + {\rm h.c.}  \right] \,, \\
    \tauYukTwoDipoleGG =& {\rm Tr}\left[  Y_d \coef{6}{q d G H}{} Y_d \coef{6}{q d G H}{} + Y_u \coef{6}{q u G H}{} Y_u \coef{6}{q u G H}{} + {\rm h.c.}  \right] \,, \\
    \tauYukTwoDipoleGGtilde =& {\rm Tr}\left[  iY_d \coef{6}{q d G H}{} Y_d \coef{6}{q d G H}{} + iY_u \coef{6}{q u G H}{} Y_u \coef{6}{q u G H}{} + {\rm h.c.}  \right] \,.
\end{align}

\subsection{Field Anomalous Dimensions}

The fermionic contributions to the wavefunction factors are
\begin{align}
	\gamma_{H} =& \gamma_{H}^{(Y)} \,, \\
	\gamma_{G} =&  \frac{4 n_g}{3} g_3^2\,, \\
	\gamma_{W} =&  \frac{4 n_g}{3} g_2^2 \,, \\
	\gamma_{B} =&  \frac{20 n_g}{9} g_1^2 \,. 
\end{align}
This comes in addition to the bosonic contributions to the wavefunction factors. We have not listed the contribution to the RGEs coming from the wavefunction factors below, but it is straightforward to add them. For the RGE for any coefficient $C$, the right-hand side should have a term $C ( \
\sum_i \gamma_i)$ where the sum is over all fields in the operator corresponding to $C$, weighted by the multiplicity of the field.

\subsection{Dimension 0}

The RGE for the cosmological constant is
\begin{align}
	\dot \Lambda &= 0 \,.
\end{align}
%

\subsection{Dimension 2}

The RGE for the Higgs mass is
\begin{align}
	\dot m_H^2 &= 0 \,.
\end{align}
%

\subsection{Dimension 4}

The RGE for the Higgs self-coupling is
\begin{align}
	\dot \lambda = & 
    m^{2}_{H} \left[ \frac{4}{3} g^{2}_{2} \kappaPauliKinSix + \kappaYukawaFourYukawaSix - 4 \kappaYukawaKinThreeSix - 2 \kappaYukawaKinudSix  \right] \,.
\end{align}
%

\subsection{Dimension 6}

The RGEs for the dimension-six coefficients in the SMEFT Lagrangian are listed below. The dimension-eight contributions are all of order $m_H^2/M^4$ in the SMEFT power counting. The dimension-six contributions agree with refs.~\cite{Jenkins:2013zja,Jenkins:2013wua,Alonso:2013hga}.

\subsubsection{$H^6$}

The RGE for the $H^6$ coupling is
\begin{align}
	\dcoef{6}{H^6}{} =& \frac{16}{3} \lambda g^{2}_{2} \kappaPauliKinSix - 4 \lambda (-\kappaYukawaFourYukawaSix + 4 \kappaYukawaKinThreeSix + 2 \kappaYukawaKinudSix)  - 4 \kappaYukawaYYYyukawaSix + m^{2}_{H} \left\{
		\left( \frac{1}{3} g^{2}_{1} \;\coef{6}{H^4 D^2}{} + \frac{2}{3} g_{1} g_{2} \;\coef{6}{WB H^2}{} \right) \kappaHypKinSix 
	\right. \nonumber \\ & \left.
		+ \left(- \frac{16}{3} g^{2}_{2} \;\coef{6}{H^4 \Box}{} + g^{2}_{2} \;\coef{6}{H^4 D^2}{} + \frac{4}{3} g_{1} g_{2} \;\coef{6}{WBH^2}{} \right) \kappaPauliKinSix
  	\right. \nonumber \\ & \left.
  + \left( 4 \;\coef{6}{H^4 \Box}{} - \frac{1}{2} \;\coef{6}{H^4 D^2}{} \right) (-\kappaYukawaFourYukawaSix + 4 \kappaYukawaKinThreeSix + 2 \kappaYukawaKinudSix)
    -  \kappaYukawaFourYukawaEight
    + 2 \kappaYukawaSixYukawaSix 
    + \frac{1}{3} g^{2}_{1} \kappaHypKinEight + \frac{1}{3} g^{2}_{2} \kappaPauliKinEight  
	  	\right. \nonumber \\ & \left.	
    + \frac{1}{3} g^{2}_{2} \kappaPauliKinEightTwo
  + \frac{1}{3} g^{2}_{2} \kappaKinSixKinSix + \frac{1}{3} (2 g^{2}_{1} - g^2_2) \kappaKinSixKinSixSUTwo + \frac{1}{12} ( -g^2_1 + g^{2}_{2} ) \kappaKinSixKinSixud
		+ \kappaYukawaKinSixKinSix - \kappaYukawaKinEight
   + \kappaYukawaSixKinSix - \kappaYukawaKinThreeEight 
   \right. \nonumber \\ & \left.
   - \kappaYukawaKinTwoEight + \kappaYukawaKinSixKinSixThree + \kappaYukawaSixKinSixThree 
   + \kappaYukawaKinSixThreeKinSixThree
   + \frac{5}{3} g_1 g_2 \tauBW + \frac{2}{3}  g_2^2 \tauWWp
   - g_2 \tauYukKinSixDipoleW
   - 2 g_1 \tauYukKinThreeSixDipoleB
   - \frac{1}{2} g_2 \tauYukKinudSixDipoleW
	\right\} \,.
\end{align}
%

\subsubsection{$H^4D^2$}

The RGEs for the $H^4D^2$ couplings are
\begin{align}
	\dcoef{6}{H^4\Box}{} =& \frac{2}{3} g^{2}_{1} \kappaHypKinSix + 2 g^{2}_{2} \kappaPauliKinSix - 2 \kappaYukawaKinSix - 6 \kappaYukawaKinThreeSix - 2 \kappaYukawaKinudSix \,,
\end{align}
\begin{align}
	\dcoef{6}{H^4 D^2}{} = & \frac{8}{3} g^{2}_{1} \kappaHypKinSix - 8 \kappaYukawaKinSix + 4 \kappaYukawaKinudSix \,.
\end{align}
%

\subsubsection{$X^2H^2$}

The RGEs for the $X^2H^2$ couplings are
\begin{align}
	\dcoef{6}{G^2H^2}{(1)}  =& - 2 g_3 \tauYukDipoleGGSix \,,
\end{align}
\begin{align}
	\dcoef{6}{G^2 H^2}{(2)}  =& 2 g_3 \tauYukDipoleGGtildeSix \,,
\end{align}
\begin{align}
	\dcoef{6}{W^2H^2}{(1)}  =& - g_2 \tauYukDipoleWWSix \,,
\end{align}
\begin{align}
	\dcoef{6}{W^2 H^2}{(2)}  =&  g_2 \tauYukDipoleWtildeWSix \,,
\end{align}
\begin{align}
	\dcoef{6}{B^2H^2}{(1)} =& - 2 g_1 \tauYukDipoleBBSix  \,,
\end{align}
\begin{align}
	\dcoef{6}{B^2 H^2}{(2)} =&  2 g_1 \tauYukDipoleBBtildeSix \,,
\end{align}
\begin{align}
	\dcoef{6}{WBH^2}{(1)}  =& - 2 g_1 \tauYukDipoleBWSix - g_2 \tauYukDipoleWBSix \,,
\end{align}
\begin{align}
	\dcoef{6}{WBH^2}{(2)}  =&  2 g_1 \tauYukDipoleBWtildeSix + g_2 \tauYukDipoleWtildeBSix \,.
\end{align}
%

\subsection{Dimension 8}

The dimension-eight RGEs for the dimension-eight coefficients in the SMEFT Lagrangian are listed below. The contributions are all of order $1/M^4$ in the SMEFT power counting.

\subsubsection{$H^8$}

The RGE for the $H^8$ coupling is
\begin{align}
	\dcoef{8}{H^8}{} =& \lambda \left( - \frac{4}{3} g^{2}_{1} \;\coef{6}{H^4D^2}{} - \frac{8}{3} g_{1} g_{2} \;\coef{6}{WB H^2}{} \right) \kappaHypKinSix 
	\nonumber \\ & 
	+ \left( - 8 g^{2}_{2} \;\coef{6}{H^6}{} + \lambda \left( \frac{64}{3} g^{2}_{2} \;\coef{6}{H^4 \Box}{} - 4 g^{2}_{2} \;\coef{6}{H^4 D^2}{} - \frac{16}{3} g_{1} g_{2} \;\coef{6}{WBH^2}{} \right) \right) \kappaPauliKinSix
	\nonumber \\ &
    + \left( 6 \;\coef{6}{H^6}{} - 16 \lambda \;\coef{6}{H^4 \Box}{} + 2 \lambda \;\coef{6}{H^4 D^2}{} \right) (-\kappaYukawaFourYukawaSix + 4 \kappaYukawaKinThreeSix + 2 \kappaYukawaKinudSix) 
    	\nonumber \\ &
     - \frac{4}{3} \lambda g^{2}_{1} \kappaHypKinEight - \frac{4}{3} \lambda g^{2}_{2} \kappaPauliKinEight - \frac{4}{3} \lambda g^{2}_{2} \kappaPauliKinEightTwo
	- \frac{4}{3} \lambda g^{2}_{2} \kappaKinSixKinSix - \frac{4}{3} \lambda ( 2 g^{2}_{1} - g^2_2 ) \kappaKinSixKinSixSUTwo + \frac{1}{3} \lambda ( g^2_1 - g^{2}_{2} ) \kappaKinSixKinSixud
     	\nonumber \\ &
          + 4 \lambda \kappaYukawaFourYukawaEight
        - 8 \lambda \kappaYukawaSixYukawaSix
        + 4 \lambda \kappaYukawaKinEight 
        + 4 \lambda \kappaYukawaKinThreeEight
        + 4 \lambda \kappaYukawaKinTwoEight
        - 4 \lambda \kappaYukawaKinSixKinSix
        - 4 \lambda \kappaYukawaKinSixKinSixThree
        - 4 \lambda \kappaYukawaKinSixThreeKinSixThree
        - 4 \lambda \kappaYukawaSixKinSix  
        \nonumber \\ &
        - 4 \lambda \kappaYukawaSixKinSixThree
        - 4 \kappaYukawaYYYyukawaEight
        + 2 \kappaYukawaYYyukawaSixYukawaSix   
    - \frac{20}{3} \lambda g_1 g_2 \tauBW - \frac{8}{3} \lambda g_2^2 \tauWWp
    + 4 \lambda g_2 \tauYukKinSixDipoleW
    + 8 \lambda g_1 \tauYukKinThreeSixDipoleB
    + 2 \lambda g_2 \tauYukKinudSixDipoleW
      \,.
\end{align}
%

\subsubsection{$H^6D^2$}

The RGEs for the $H^6D^2$ couplings are
\begin{align}
	\dcoef{8}{H^6D^2}{(1)} =& \left( 2 g^{2}_{1} \;\coef{6}{H^4 D^2}{} + \frac{16}{3} g_{1} g_{2} \;\coef{6}{WB H^2}{} \right) \kappaHypKinSix 
	\nonumber \\ &
	+ \left( - \frac{32}{3} g^{2}_{2} \;\coef{6}{H^4 \Box}{} + \frac{2}{3} g^{2}_{2} \;\coef{6}{H^4 D^2}{} + 8 g_{1} g_{2} \;\coef{6}{WBH^2}{} \right) \kappaPauliKinSix
	\nonumber \\ &
    + \left( 8 \;\coef{6}{H^4 \Box}{} + \;\coef{6}{H^4 D^2}{} \right)(-\kappaYukawaFourYukawaSix + 4 \kappaYukawaKinThreeSix + 2 \kappaYukawaKinudSix) 
    \nonumber \\ & 
    + 2 g^{2}_{1} \kappaHypKinEight + \frac{10}{3} g^{2}_{2} \kappaPauliKinEight + 2 g^{2}_{2} \kappaPauliKinEightTwo
	+ \frac{8}{3} g^{2}_{2} \kappaKinSixKinSix +\left(4 g^{2}_{1} - \frac{10}{3} g^2_2 \right)\kappaKinSixKinSixSUTwo + \frac{1}{2} ( - g^2_1 + 2 g^{2}_{2} ) \kappaKinSixKinSixud 
    \nonumber \\ &
    + 2 \kappaYukawaSixYukawaSix - 6 \kappaYukawaKinEight - 10 \kappaYukawaKinThreeEight - 2 \kappaYukawaKinudEight - 6 \kappaYukawaKinTwoEight + 6 \kappaYukawaKinSixKinSix + 6 \kappaYukawaKinSixKinSixThree + 10 \kappaYukawaKinSixThreeKinSixThree + 6 \kappaYukawaSixKinSix + 10 \kappaYukawaSixKinSixThree
    \nonumber \\ &
    + 2 \kappaYukawaSixKinSixud - \kappaYukawaKinSixudKinSixud + 4 \kappaYukawaKinSixThreeKinSixud
    + \frac{32}{3} g_1 g_2 \tauBW + \frac{20}{3} g_2^2 \tauWWp 
    - 8 g_2 \tauYukKinSixDipoleW
    - 12 g_1 \tauYukKinThreeSixDipoleB
    - 6 g_2 \tauYukKinudSixDipoleW
    \,,
\end{align}
\begin{align}
	\dcoef{8}{H^6D^2}{(2)} =& \left( \frac{4}{3} g^{2}_{1} \;\coef{6}{H^4 D^2}{} + \frac{4}{3} g_{1} g_{2} \;\coef{6}{WB H^2}{} \right) \kappaHypKinSix 
	+ \left( - \frac{4}{3} g^{2}_{2} \;\coef{6}{H^4 D^2}{} + \frac{16}{3} g_{1} g_{2} \;\coef{6}{WBH^2}{} \right) \kappaPauliKinSix
	\nonumber \\ &
    + \left( 2 \;\coef{6}{H^4 D^2}{} \right) (-\kappaYukawaFourYukawaSix + 4 \kappaYukawaKinThreeSix + 2 \kappaYukawaKinudSix) 
    + \frac{4}{3} g^{2}_{1} \kappaHypKinEight + \frac{4}{3} g^{2}_{2} \kappaPauliKinEightTwo
	+ \frac{2}{3} g^{2}_{2} \kappaKinSixKinSix + \frac{8}{3} g^{2}_{1} \kappaKinSixKinSixSUTwo 
     \nonumber \\ & 
    - \frac{1}{6} (2 g^2_1 + g^{2}_{2}) \kappaKinSixKinSixud 
   - 4 \kappaYukawaKinEight + 2 \kappaYukawaKinudEight - 4 \kappaYukawaKinTwoEight  + 4 \kappaYukawaKinSixKinSix  + 4 \kappaYukawaKinSixKinSixThree  + 4 \kappaYukawaSixKinSix 
   \nonumber \\ &
   - 2 \kappaYukawaSixKinSixud + \kappaYukawaKinSixudKinSixud - 4 \kappaYukawaKinSixThreeKinSixud   
   + 6 g_1 g_2 \tauBW 
   - 2 g_2 \tauYukKinSixDipoleW
   - 8 g_1 \tauYukKinThreeSixDipoleB
   + g_2 \tauYukKinudSixDipoleW
    \,.
\end{align}
%

\subsubsection{$H^4D^4$}

The RGEs for the $H^4D^4$ couplings are
\begin{align}
	\dcoef{8}{H^4 D^4}{(1)}  =& \frac{8}{3} \kappaKinSixKinSix - \frac{8}{3} \kappaKinSixKinSixSUTwo - \frac{4}{3} \kappaKinSixKinSixud \,,
\end{align}
\begin{align}
	\dcoef{8}{H^4 D^4}{(2)} =& - \frac{8}{3} \kappaKinSixKinSix - \frac{8}{3} \kappaKinSixKinSixSUTwo \,,
\end{align}
\begin{align}
	\dcoef{8}{H^4 D^4}{(3)}  =&  \frac{16}{3} \kappaKinSixKinSixSUTwo + \frac{4}{3} \kappaKinSixKinSixud \,.
\end{align}
%

\subsubsection{$X^2H^4$}

The RGEs for the $X^2H^4$ coefficients are
\begin{align}
	\dcoef{8}{G^2H^4}{(1)} =& \left( - \frac{8}{3} g^{2}_{2} \;\coef{6}{G^2 H^2}{} \right) \kappaPauliKinSix 
    + \left( 2 \;\coef{6}{G^2 H^2}{} \right)(-\kappaYukawaFourYukawaSix + 4 \kappaYukawaKinThreeSix + 2 \kappaYukawaKinudSix) 
    - 2 g_3 \tauYukDipoleGGEight
    + 2 \tauYukTwoDipoleGG
    \,,
\end{align}
\begin{align}
	\dcoef{8}{G^2H^4}{(2)} =& 2 g_3 \tauYukDipoleGGtildeEight - 2 \tauYukTwoDipoleGGtilde
    \,,
\end{align}
\begin{align}
	\dcoef{8}{W^2H^4}{(1)}  =& \left( - \frac{2}{3} g_{1} g_{2} \;\coef{6}{WB H^2}{} \right) \kappaHypKinSix 
    + \left( 2 \;\coef{6}{W^2 H^2}{} \right)(-\kappaYukawaFourYukawaSix + 4 \kappaYukawaKinThreeSix + 2 \kappaYukawaKinudSix)  
    - \frac{1}{3} g^{2}_{2} \kappaKinSixKinSix + g^{2}_{2} \kappaKinSixKinSixSUTwo 
    \nonumber \\ &
    - \frac{1}{6} g^{2}_{2} \kappaKinSixKinSixud 
    - \frac{1}{3} (g_1 g_2 \tauBW - 2 g_2^2 \tauWW + 4 g_2^2 \tauWWp )
    - g_2 \tauYukDipoleWWEight 
    + g_2 \tauYukKinSixDipoleW
    - 2 g_2 \tauYukKinThreeSixDipoleW
    + \frac{1}{2} g_2 \tauYukKinudSixDipoleW
    + 8 \tauYukTwoDipoleWWtwo
    \,,
\end{align}
\begin{align}
	\dcoef{8}{W^2H^4}{(2)}  =& 
    - \frac{1}{3} g_1 g_2 \tauBWp 
    + g_2 \tauYukDipoleWtildeWEight
    - g_2 \tauYukKinSixDipoleWtilde
    + 2 g_2 \tauYukKinThreeSixDipoleWtilde
    - \frac{1}{2} g_2 \tauYukKinudSixDipoleWtilde
    - 8 \tauYukTwoDipoleWWtildetwo
    \,,
\end{align}
\begin{align}
	\dcoef{8}{W^2H^4}{(3)} =& \left( \frac{2}{3} g_{1} g_{2} \;\coef{6}{WB H^2}{} \right) \kappaHypKinSix 
    + \frac{1}{3} g_1 g_2 \tauBW - g_2 \tauYukDipoleWWtwoEight
    - g_2 \tauYukKinSixDipoleW
    + \frac{1}{2} g_2 \tauYukKinudSixDipoleW
    + 4 \tauYukTwoDipoleWW
    - 8 \tauYukTwoDipoleWWtwo
    \,,
\end{align}
\begin{align}
	\dcoef{8}{W^2H^4}{(4)}  =& 
    \frac{1}{3} g_1 g_2 \tauBWp 
    + g_2 \tauYukDipoleWtildeWtwoEight
    + g_2 \tauYukKinSixDipoleWtilde
    - \frac{1}{2} g_2 \tauYukKinudSixDipoleWtilde
    - 4 \tauYukTwoDipoleWWtilde
    + 8 \tauYukTwoDipoleWWtildetwo
    \,,
\end{align}
\begin{align}
	\dcoef{8}{B^2H^4}{(1)} =& \left( \frac{8}{3} g^{2}_{1} \;\coef{6}{B^2 H^2}{} \right) \kappaHypKinSix
	+ \left( - \frac{8}{3} g^{2}_{2} \;\coef{6}{B^2 H^2}{} - \frac{4}{3} g_{1} g_{2} \;\coef{6}{WBH^2}{} \right) \kappaPauliKinSix 
     \nonumber \\ &
    + \left( 2 \;\coef{6}{B^2 H^2}{} \right)(-\kappaYukawaFourYukawaSix + 4 \kappaYukawaKinThreeSix + 2 \kappaYukawaKinudSix)  
	+ \frac{1}{3} g^{2}_{1} \kappaKinSixKinSix - g^{2}_{1} \kappaKinSixKinSixSUTwo + \frac{1}{6} g^2_1 \kappaKinSixKinSixud
     \nonumber \\ &
    + \frac{2}{3} (g_1^2 \tauBB - g_1 g_2 \tauBW )
    - 2 g_{1} \tauYukDipoleBBEight
    - 2 g_1 \tauYukKinSixDipoleB + 4 g_1 \tauYukKinThreeSixDipoleB
    + 4 \tauYukTwoDipoleBB
    \,,
\end{align}
\begin{align}
	\dcoef{8}{B^2H^4}{(2)} =& \frac{2}{3} g_1 g_2 \tauBWp + 2 g_{1} \tauYukDipoleBBtildeEight
    + 2 g_1 \tauYukKinSixDipoleBtilde
    - 4 g_1 \tauYukKinThreeSixDipoleBtilde
    - 4 \tauYukTwoDipoleBBtilde
    \,,
\end{align}
\begin{align}
	\dcoef{8}{WBH^4}{(1)} =& \left( \frac{8}{3} g_{1} g_{2} \;\coef{6}{B^2 H^2}{} \right) \kappaHypKinSix 
	+ \left( \frac{8}{3} g_{1} g_{2} \;\coef{6}{W^2 H^2}{} - 4 g^{2}_{2} \;\coef{6}{WBH^2}{} \right) \kappaPauliKinSix
	\nonumber \\ &
    + \left( 2 \;\coef{6}{WB H^2}{} \right)(-\kappaYukawaFourYukawaSix + 4 \kappaYukawaKinThreeSix + 2 \kappaYukawaKinudSix) 
    \nonumber \\ &
    + \frac{2}{3} (g_1 g_2 \tauBB - g_2^2 \tauBW + g_1 g_2 \tauWW - 2 g_1 g_2 \tauWWp )
    - 2 g_{1} \tauYukDipoleBWEight - g_2 \tauYukDipoleWBEight
    - 2 g_2 \tauYukKinSixDipoleB
    + 2 g_2 \tauYukKinThreeSixDipoleB
    \nonumber \\ & 
    - 2 g_1 \tauYukKinThreeSixDipoleW
    + 8 \tauYukTwoDipoleBW
    + g_1 \tauYukKinudSixDipoleW
    \,,
\end{align}
\begin{align}
	\dcoef{8}{WBH^4}{(2)} =&  \frac{2}{3} g^2_2 \tauBWp + 2 g_{1} \tauYukDipoleBWtildeEight + g_2 \tauYukDipoleWtildeBEight
    + 2 g_2 \tauYukKinSixDipoleBtilde
    - 2 g_2 \tauYukKinThreeSixDipoleBtilde
    + 2 g_1 \tauYukKinThreeSixDipoleWtilde
    - 8 \tauYukTwoDipoleBWtilde
    - g_1 \tauYukKinudSixDipoleWtilde
    \,.
\end{align}
%

\subsubsection{$XH^4D^2$}

The RGEs for the $XH^4D^2$ couplings are
\begin{align}
	\dcoef{8}{BH^4D^2}{(1)} =& \left( - \frac{32}{3} g_{1} \;\coef{6}{B^2 H^2}{} \right) \kappaHypKinSix 
	+ \left( \frac{48}{3} g_{2} \;\coef{6}{WBH^2}{} \right) \kappaPauliKinSix
	- \frac{8}{3} g_{1} \kappaKinSixKinSix + 8 g_{1} \kappaKinSixKinSixSUTwo -  \frac{4}{3} g_{1} \kappaKinSixKinSixud 
    \nonumber \\ &
    - \frac{8}{3} (g_1 \tauBB - 3 g_2 \tauBW )
    + 8 \tauYukKinSixDipoleB
    - 32 \tauYukKinThreeSixDipoleB
 \,,
\end{align}
\begin{align}
	\dcoef{8}{BH^4D^2}{(2)} =& - 8 g_2 \tauBWp - 8 \tauYukKinSixDipoleBtilde + 32 \tauYukKinThreeSixDipoleBtilde
    \,,
\end{align}
\begin{align}
	\dcoef{8}{WH^4D^2}{(1)} =& \left( \frac{16}{3} g_{1} \;\coef{6}{B^2 H^2}{} \right) \kappaHypKinSix 
	+ \left( - \frac{32}{3} g_{2} \;\coef{6}{W^2 H^2}{} \right) \kappaPauliKinSix
	+ \frac{8}{3} g_{2} \kappaKinSixKinSix - 8 g_{2} \kappaKinSixKinSixSUTwo + \frac{4}{3} g_{2} \kappaKinSixKinSixud 
    \nonumber \\ &
    + \frac{8}{3} ( g_1 \tauBW -  g_2 \tauWW + 4 g_2 \tauWWp )
    - 8 \tauYukKinSixDipoleW
    + 8 \tauYukKinThreeSixDipoleW
    - 8 \tauYukKinudSixDipoleW
    \,,
\end{align}
\begin{align}
	\dcoef{8}{WH^4D^2}{(2)} =&  
    - \frac{8}{3} g_1 \tauBWp 
    + 8 \tauYukKinSixDipoleWtilde
    - 8 \tauYukKinThreeSixDipoleWtilde
    + 8 \tauYukKinudSixDipoleWtilde
    \,,
\end{align}
\begin{align}
	\dcoef{8}{WH^4D^2}{(3)} =& 
    \frac{8}{3} g_1 \tauBWp 
    + 8 \tauYukKinThreeSixDipoleWeps 
    + 4 \tauYukKinudSixDipoleWeps \,,
\end{align}
\begin{align}
	\dcoef{8}{WH^4D^2}{(4)} =& 
    - \frac{8}{3} g_1 \tauBW + \frac{8}{3} g_2 \tauWW  - \frac{8}{3} g_2 \tauWWp 
    + 8 \tauYukKinThreeSixDipoleWtildeeps 
    + 4 \tauYukKinudSixDipoleWtildeeps \,.
\end{align}
%

\subsubsection{$X^2H^2D^2$}

The RGEs for the $X^2H^2D^2$ couplings are
\begin{align}
	\dcoef{8}{G^2H^2D^2}{(1)} =& \frac{16}{3} \tauGG  \,,
\end{align}
\begin{align}
	\dcoef{8}{G^2H^2D^2}{(2)} =& - \frac{4}{3} \tauGG \,, 
\end{align}
\begin{align}
	\dcoef{8}{W^2H^2D^2}{(1)} =& \frac{16}{3} \tauWW \,,
\end{align}
\begin{align}
	\dcoef{8}{W^2H^2D^2}{(2)} =& - \frac{4}{3} \tauWW  \,, 
\end{align}
\begin{align}
	\dcoef{8}{W^2H^2D^2}{(4)} =& \frac{16}{3} \tauWWp \,,
\end{align}
\begin{align}
	\dcoef{8}{W^2H^2D^2}{(5)} =& \frac{8}{3} \tauWW \,,
\end{align}
\begin{align}
	\dcoef{8}{WBH^2D^2}{(1)} =& - \frac{8}{3} \tauBW \,,
\end{align}
\begin{align}
	\dcoef{8}{WBH^2D^2}{(2)} =& -\frac{4}{3} \tauBWpp \,,
\end{align}
\begin{align}
	\dcoef{8}{WBH^2D^2}{(3)} =& \frac{16}{3} \tauBWp \,,
\end{align}
\begin{align}
	\dcoef{8}{WBH^2D^2}{(4)} =& \frac{16}{3} \tauBW \,,
\end{align}
\begin{align}
	\dcoef{8}{WBH^2D^2}{(6)} =& \frac{8}{3} \tauBWpp \,,
\end{align}
\begin{align}
	\dcoef{8}{B^2H^2D^2}{(1)} =& \frac{16}{3} \tauBB \,,
\end{align}
\begin{align}
	\dcoef{8}{B^2H^2D^2}{(2)} =& - \frac{4}{3} \tauBB \,. 
\end{align}

\section{Self-duality relations for the dipole}
\label{sec:self-dual}

We use the convention $\epsilon_{0123}=+1$.
The duality relation $\epsilon_{\mu\nu\rho\sigma}\sigma^{\rho\sigma}=2i \sigma_{\mu\nu}\gamma_5$ gives
\begin{align}
   & \frac{i}2 \epsilon_{\mu\nu\rho\sigma}\sigma^{\rho\sigma} P_R =-\sigma_{\mu\nu}P_R \,, &
   & \frac{i}2 \epsilon_{\mu\nu\rho\sigma}\sigma^{\rho\sigma} P_L =\sigma_{\mu\nu}P_L\,.
   \label{C.1}
\end{align}
The dual of a tensor $T^{\mu \nu}$ is defined by
\begin{align}
    \widetilde T_{\mu\nu} &\equiv \frac12 \epsilon_{\mu\nu\rho\sigma} T^{\rho\sigma},
    & \widetilde{\widetilde{T}}_{\mu\nu} &= -T_{\mu\nu} \,.
\end{align}
A tensor will be called selfdual if $\tilde T^{\mu\nu} = i T^{\mu\nu}$, and anti-selfdual if $\tilde T^{\mu\nu} = -i T^{\mu\nu}$. For example, the selfdual part of the electromagnetic field-strength tensor is $F_{\mu \nu} - i \widetilde F_{\mu \nu}$, and the anti-selfdual part is $F_{\mu \nu} + i \widetilde F_{\mu \nu}$.
If $T^{\mu \nu}$ is selfdual, $T^{\dagger \mu \nu}$ is anti-selfdual. Eq.~\eqref{C.1} relates the chirality of fermion dipole bilinears to their duality property,
\begin{align}
    \overline \psi\, T^{\mu \nu} \sigma_{\mu \nu} P_R \psi
& = -i \overline \psi\, \widetilde T^{\mu \nu} \sigma_{\mu \nu} P_R \psi,
\end{align}
so  $ \overline \psi\, T^{\mu \nu} \sigma_{\mu \nu} P_R \psi$ vanishes if $T^{\mu \nu}$ is anti-selfdual, i.e., $T^{\mu \nu}$ can be chosen to be selfdual.
One useful implication of this self-duality is that the chiral projectors can be dropped for tensor bilinears if $T^{\mu \nu}$ is selfdual,
\begin{align}
    \bar \psi  \sigma_{\mu\nu}T^{\mu\nu} \psi &= \bar \psi  \sigma_{\mu\nu}T^{\mu\nu} P_R \psi \,, &
     \bar \psi  \sigma_{\mu\nu}T^{\mu\nu\dag} \psi &= \bar \psi \sigma_{\mu\nu}T^{\mu\nu\dag} P_L \psi\,.
\end{align}
There are a few relations that can be derived from this self-duality condition,
\begin{align}
    & T^{\mu\nu}T_{\mu\nu}^\dag =0 \,, \\
    &  {T_\mu} ^\alpha T_{\nu \alpha}^\dag  = {T_\nu}^\alpha T_{\mu \alpha}^\dag \,, \\
    & {T_\mu} ^\alpha T_{\nu \alpha} = \frac12 g_{\mu\nu} T_{\alpha \beta}T^{\alpha \beta} - {T_\nu} ^\alpha T_{\mu \alpha} \,, \\
    &4\, T^{\mu\nu} T_{\nu \alpha}^\dag T^{\alpha\beta} T_{\beta \mu }^\dag = T_{\mu \nu} T_{\alpha \beta}^\dag T^{\mu\nu}T^{\dag \alpha \beta }\,.
\end{align}
These relations have been used to derive \cref{eq:oneloopdivcompact}.

\section{Majorana masses and dipoles} \label{app:Majorana}

If Majorana mass or dipole terms are present in the Lagrangian, they can be included in the formalism by promoting the field in eq.~\eqref{eq:LRfields} to
\begin{align}\label{eq:LRfieldswM}
    \chi = \begin{pmatrix}
        \chi_L \\
        (\chi_R)^c \\
        (\chi_L)^c \\
        \chi_R
    \end{pmatrix} \,, 
\end{align}
where $(\chi_{R})^c = C ( \overline{ \chi_{R}})^\intercal$ is left-handed, with $C = i \gamma^2 \gamma^0$ the charge conjugation matrix. In the SMEFT, the fields in \cref{eq:LRfieldswM} are $\chi_L=\{\ell, q\}$, $(\chi_R)^c=\{e^c,u^c, d^c\}$, $(\chi_L)^c=\{ \ell^c, q^c \}$, $\chi_R=\{e, u, d\}$.
In this new basis where the degrees of freedom are essentially doubled, the metric in the Lagrangian eq.~\eqref{eq:d2L-fermion} takes the $4 \times 4$ block form
\begin{equation}\label{eq:kdefsM}
    k= \frac12 \begin{pmatrix}
        \kappa & 0 \\
        0 &\kappa^\intercal
    \end{pmatrix}\,, 
    \qquad
    \kappa =  \begin{pmatrix}
        \kappa_L & 0 \\
        0 &\kappa_R^* 
    \end{pmatrix}\,.
\end{equation}
The field strength has the block form 
\begin{equation} \label{eq:fsdefM}
    \cY_{\mu\nu}= \begin{pmatrix}
        Y_L^{\mu\nu} &0&0&0 \\
        0 & Y_R^{\mu\nu\ast} &0&0 \\
       0 &0& Y_L^{\mu\nu\ast} &0 \\
        0 &0&0&Y_R^{\mu\nu} 
    \end{pmatrix} \,,
\end{equation}
with $Y_{L(R)}^{\mu\nu}=[\mathcal D_{L(R)}^\mu, \mathcal D_{L(R)}^\nu]$. The field strength tensors have the same anti-hermitian property as the gauge field, i.e., $\kappa_L^{-1}Y_L^{\mu\nu\dag} = - Y_L^{\mu\nu}\kappa_L^{-1}$ and similarly for $Y_{R}^{\mu\nu}$.
The scalar $\mathcal M$ and tensor $\mathcal{T}^{\mu\nu}$ objects retain the same global structure as in eq.~\eqref{eq:MTdefs} where $M$ and $T^{\mu \nu}$ are now $2 \times 2$ submatrices
\begin{equation}\label{eq:MTdefsM}
    M= \frac12\begin{pmatrix}
        \mu_L & m_D \\
        m_D^\intercal & \mu_R^*
    \end{pmatrix}\,,
    \qquad
    T^{\mu\nu}= \frac12\begin{pmatrix}
        t^{\mu\nu}_L & t^{\mu\nu}_D \\
        -t^{\mu\nu \intercal}_D & -t^{\mu\nu*}_R \\
    \end{pmatrix}\,,
\end{equation}
where we have defined $\mu_{L(R)}$ as the symmetric Majorana mass matrices and $t^{\mu\nu}_{L(R)}$ as the antisymmetric Majorana dipole matrices for the left-(right-)handed fields. The Dirac mass matrix is denoted as $m_D$ and the Dirac dipole matrix as $t^{\mu\nu}_D$. The one-loop divergence formula in \cref{eq:oneloopdivcompact} can be used with the above modifications, with the prefactor $1/(32 \pi^2 \epsilon)$ replaced by $1/(64 \pi^2 \epsilon)$, since each term in the trace is now counted twice, because of the doubled set of fields.
%


\bibliographystyle{JHEP}
\bibliography{bibliographyManifold.bib}

\providecommand{\href}[2]{#2}\begingroup\raggedright\begin{thebibliography}{10}

\bibitem{Chisholm:1961tha}
J.~S.~R. Chisholm, \emph{{Change of variables in quantum field theories}},
  \href{https://doi.org/10.1016/0029-5582(61)90106-7}{\emph{Nucl. Phys.}
  {\bfseries 26} (1961) 469}.

\bibitem{Kamefuchi:1961sb}
S.~Kamefuchi, L.~O'Raifeartaigh and A.~Salam, \emph{{Change of variables and
  equivalence theorems in quantum field theories}},
  \href{https://doi.org/10.1016/0029-5582(61)90056-6}{\emph{Nucl. Phys.}
  {\bfseries 28} (1961) 529}.

\bibitem{Politzer:1980me}
H.~D. Politzer, \emph{{Power Corrections at Short Distances}},
  \href{https://doi.org/10.1016/0550-3213(80)90172-8}{\emph{Nucl. Phys. B}
  {\bfseries 172} (1980) 349}.

\bibitem{Arzt:1993gz}
C.~Arzt, \emph{{Reduced effective Lagrangians}},
  \href{https://doi.org/10.1016/0370-2693(94)01419-D}{\emph{Phys. Lett. B}
  {\bfseries 342} (1995) 189}
  [\href{https://arxiv.org/abs/hep-ph/9304230}{{\ttfamily hep-ph/9304230}}].

\bibitem{Meetz:1969as}
K.~Meetz, \emph{{Realization of chiral symmetry in a curved isospin space}},
  \href{https://doi.org/10.1063/1.1664881}{\emph{J. Math. Phys.} {\bfseries 10}
  (1969) 589}.

\bibitem{Honerkamp:1971xtx}
J.~Honerkamp and K.~Meetz, \emph{{Chiral-invariant perturbation theory}},
  \href{https://doi.org/10.1103/PhysRevD.3.1996}{\emph{Phys. Rev. D} {\bfseries
  3} (1971) 1996}.

\bibitem{Honerkamp:1971sh}
J.~Honerkamp, \emph{{Chiral multiloops}},
  \href{https://doi.org/10.1016/0550-3213(72)90299-4}{\emph{Nucl. Phys. B}
  {\bfseries 36} (1972) 130}.

\bibitem{Volkov:1970aa}
D.~V. Volkov, \emph{Phenomenological lagrangians}, {\emph{Sov. J. Particles
  Nucl.} {\bfseries 4} (1973) 1}.

\bibitem{Cheung:2021yog}
C.~Cheung, A.~Helset and J.~Parra-Martinez, \emph{{Geometric soft theorems}},
  \href{https://doi.org/10.1007/JHEP04(2022)011}{\emph{JHEP} {\bfseries 04}
  (2022) 011} [\href{https://arxiv.org/abs/2111.03045}{{\ttfamily
  2111.03045}}].

\bibitem{Alonso:2015fsp}
R.~Alonso, E.~E. Jenkins and A.~V. Manohar, \emph{{A Geometric Formulation of
  Higgs Effective Field Theory: Measuring the Curvature of Scalar Field
  Space}}, \href{https://doi.org/10.1016/j.physletb.2016.01.041}{\emph{Phys.
  Lett. B} {\bfseries 754} (2016) 335}
  [\href{https://arxiv.org/abs/1511.00724}{{\ttfamily 1511.00724}}].

\bibitem{Alonso:2016oah}
R.~Alonso, E.~E. Jenkins and A.~V. Manohar, \emph{{Geometry of the Scalar
  Sector}}, \href{https://doi.org/10.1007/JHEP08(2016)101}{\emph{JHEP}
  {\bfseries 08} (2016) 101}
  [\href{https://arxiv.org/abs/1605.03602}{{\ttfamily 1605.03602}}].

\bibitem{Alonso:2017tdy}
R.~Alonso, K.~Kanshin and S.~Saa, \emph{{Renormalization group evolution of
  Higgs effective field theory}},
  \href{https://doi.org/10.1103/PhysRevD.97.035010}{\emph{Phys. Rev. D}
  {\bfseries 97} (2018) 035010}
  [\href{https://arxiv.org/abs/1710.06848}{{\ttfamily 1710.06848}}].

\bibitem{Helset:2018fgq}
A.~Helset, M.~Paraskevas and M.~Trott, \emph{{Gauge fixing the Standard Model
  Effective Field Theory}},
  \href{https://doi.org/10.1103/PhysRevLett.120.251801}{\emph{Phys. Rev. Lett.}
  {\bfseries 120} (2018) 251801}
  [\href{https://arxiv.org/abs/1803.08001}{{\ttfamily 1803.08001}}].

\bibitem{Helset:2020yio}
A.~Helset, A.~Martin and M.~Trott, \emph{{The Geometric Standard Model
  Effective Field Theory}},
  \href{https://doi.org/10.1007/JHEP03(2020)163}{\emph{JHEP} {\bfseries 03}
  (2020) 163} [\href{https://arxiv.org/abs/2001.01453}{{\ttfamily
  2001.01453}}].

\bibitem{Hays:2020scx}
C.~Hays, A.~Helset, A.~Martin and M.~Trott, \emph{{Exact SMEFT formulation and
  expansion to $\mathcal{O}(v^4/\Lambda^4)$}},
  \href{https://doi.org/10.1007/JHEP11(2020)087}{\emph{JHEP} {\bfseries 11}
  (2020) 087} [\href{https://arxiv.org/abs/2007.00565}{{\ttfamily
  2007.00565}}].

\bibitem{Cohen:2020xca}
T.~Cohen, N.~Craig, X.~Lu and D.~Sutherland, \emph{{Is SMEFT Enough?}},
  \href{https://doi.org/10.1007/JHEP03(2021)237}{\emph{JHEP} {\bfseries 03}
  (2021) 237} [\href{https://arxiv.org/abs/2008.08597}{{\ttfamily
  2008.08597}}].

\bibitem{Corbett:2021eux}
T.~Corbett, A.~Helset, A.~Martin and M.~Trott, \emph{{EWPD in the SMEFT to
  dimension eight}}, \href{https://doi.org/10.1007/JHEP06(2021)076}{\emph{JHEP}
  {\bfseries 06} (2021) 076}
  [\href{https://arxiv.org/abs/2102.02819}{{\ttfamily 2102.02819}}].

\bibitem{Corbett:2021cil}
T.~Corbett, A.~Martin and M.~Trott, \emph{{Consistent higher order $ \sigma
  \left(\mathcal{GG}\to h\right) $, $ \Gamma \left(h\to \mathcal{GG}\right) $
  and \ensuremath{\Gamma}(h \textrightarrow{}
  \ensuremath{\gamma}\ensuremath{\gamma}) in geoSMEFT}},
  \href{https://doi.org/10.1007/JHEP12(2021)147}{\emph{JHEP} {\bfseries 12}
  (2021) 147} [\href{https://arxiv.org/abs/2107.07470}{{\ttfamily
  2107.07470}}].

\bibitem{Cohen:2021ucp}
T.~Cohen, N.~Craig, X.~Lu and D.~Sutherland, \emph{{Unitarity violation and the
  geometry of Higgs EFTs}},
  \href{https://doi.org/10.1007/JHEP12(2021)003}{\emph{JHEP} {\bfseries 12}
  (2021) 003} [\href{https://arxiv.org/abs/2108.03240}{{\ttfamily
  2108.03240}}].

\bibitem{Martin:2023fad}
A.~Martin and M.~Trott, \emph{{More accurate $\sigma(\mathcal{G}
  \,\mathcal{G}\rightarrow h)$, $\Gamma(h \rightarrow \mathcal{G}
  \,\mathcal{G}, \mathcal{A} \mathcal{A}, \bar{\Psi} {\Psi})$ and Higgs width
  results via the geoSMEFT}},
  \href{https://arxiv.org/abs/2305.05879}{{\ttfamily 2305.05879}}.

\bibitem{Gattus:2023gep}
V.~Gattus and A.~Pilaftsis, \emph{{Minimal Supergeometric Quantum Field
  Theories}},  \href{https://arxiv.org/abs/2307.01126}{{\ttfamily 2307.01126}}.

\bibitem{Finn:2020nvn}
K.~Finn, S.~Karamitsos and A.~Pilaftsis, \emph{{Frame covariant formalism for
  fermionic theories}},
  \href{https://doi.org/10.1140/epjc/s10052-021-09360-w}{\emph{Eur. Phys. J. C}
  {\bfseries 81} (2021) 572}
  [\href{https://arxiv.org/abs/2006.05831}{{\ttfamily 2006.05831}}].

\bibitem{Cheung:2022vnd}
C.~Cheung, A.~Helset and J.~Parra-Martinez, \emph{{Geometry-kinematics
  duality}}, \href{https://doi.org/10.1103/PhysRevD.106.045016}{\emph{Phys.
  Rev. D} {\bfseries 106} (2022) 045016}
  [\href{https://arxiv.org/abs/2202.06972}{{\ttfamily 2202.06972}}].

\bibitem{Cohen:2022uuw}
T.~Cohen, N.~Craig, X.~Lu and D.~Sutherland, \emph{{On-Shell Covariance of
  Quantum Field Theory Amplitudes}},
  \href{https://doi.org/10.1103/PhysRevLett.130.041603}{\emph{Phys. Rev. Lett.}
  {\bfseries 130} (2023) 041603}
  [\href{https://arxiv.org/abs/2202.06965}{{\ttfamily 2202.06965}}].

\bibitem{Helset:2022tlf}
A.~Helset, E.~E. Jenkins and A.~V. Manohar, \emph{{Geometry in scattering
  amplitudes}}, \href{https://doi.org/10.1103/PhysRevD.106.116018}{\emph{Phys.
  Rev. D} {\bfseries 106} (2022) 116018}
  [\href{https://arxiv.org/abs/2210.08000}{{\ttfamily 2210.08000}}].

\bibitem{Craig:2023wni}
N.~Craig, Y.-T. Lee, X.~Lu and D.~Sutherland, \emph{{Effective Field Theories
  as Lagrange Spaces}},  \href{https://arxiv.org/abs/2305.09722}{{\ttfamily
  2305.09722}}.

\bibitem{Helset:2022pde}
A.~Helset, E.~E. Jenkins and A.~V. Manohar, \emph{{Renormalization of the
  Standard Model Effective Field Theory from geometry}},
  \href{https://doi.org/10.1007/JHEP02(2023)063}{\emph{JHEP} {\bfseries 02}
  (2023) 063} [\href{https://arxiv.org/abs/2212.03253}{{\ttfamily
  2212.03253}}].

\bibitem{Jenkins:2013zja}
E.~E. Jenkins, A.~V. Manohar and M.~Trott, \emph{{Renormalization Group
  Evolution of the Standard Model Dimension Six Operators I: Formalism and
  $\lambda$ Dependence}},
  \href{https://doi.org/10.1007/JHEP10(2013)087}{\emph{JHEP} {\bfseries 10}
  (2013) 087} [\href{https://arxiv.org/abs/1308.2627}{{\ttfamily 1308.2627}}].

\bibitem{Jenkins:2013wua}
E.~E. Jenkins, A.~V. Manohar and M.~Trott, \emph{{Renormalization Group
  Evolution of the Standard Model Dimension Six Operators II: Yukawa
  Dependence}}, \href{https://doi.org/10.1007/JHEP01(2014)035}{\emph{JHEP}
  {\bfseries 01} (2014) 035} [\href{https://arxiv.org/abs/1310.4838}{{\ttfamily
  1310.4838}}].

\bibitem{Alonso:2013hga}
R.~Alonso, E.~E. Jenkins, A.~V. Manohar and M.~Trott, \emph{{Renormalization
  Group Evolution of the Standard Model Dimension Six Operators III: Gauge
  Coupling Dependence and Phenomenology}},
  \href{https://doi.org/10.1007/JHEP04(2014)159}{\emph{JHEP} {\bfseries 04}
  (2014) 159} [\href{https://arxiv.org/abs/1312.2014}{{\ttfamily 1312.2014}}].

\bibitem{Chala:2021pll}
M.~Chala, G.~Guedes, M.~Ramos and J.~Santiago, \emph{{Towards the
  renormalisation of the Standard Model effective field theory to dimension
  eight: Bosonic interactions I}},
  \href{https://doi.org/10.21468/SciPostPhys.11.3.065}{\emph{SciPost Phys.}
  {\bfseries 11} (2021) 065}
  [\href{https://arxiv.org/abs/2106.05291v3}{{\ttfamily 2106.05291v3}}].

\bibitem{DasBakshi:2022mwk}
S.~Das~Bakshi, M.~Chala, A.~D\'\i{}az-Carmona and G.~Guedes, \emph{{Towards the
  renormalisation of the Standard Model effective field theory to dimension
  eight: bosonic interactions II}},
  \href{https://doi.org/10.1140/epjp/s13360-022-03194-5}{\emph{Eur. Phys. J.
  Plus} {\bfseries 137} (2022) 973}
  [\href{https://arxiv.org/abs/2205.03301}{{\ttfamily 2205.03301}}].

\bibitem{AccettulliHuber:2021uoa}
M.~Accettulli~Huber and S.~De~Angelis, \emph{{Standard Model EFTs via on-shell
  methods}}, \href{https://doi.org/10.1007/JHEP11(2021)221}{\emph{JHEP}
  {\bfseries 11} (2021) 221}
  [\href{https://arxiv.org/abs/2108.03669}{{\ttfamily 2108.03669}}].

\bibitem{DasBakshi:2023htx}
S.~Das~Bakshi and A.~D\'\i{}az-Carmona, \emph{{Renormalisation of SMEFT bosonic
  interactions up to dimension eight by LNV operators}},
  \href{https://doi.org/10.1007/JHEP06(2023)123}{\emph{JHEP} {\bfseries 06}
  (2023) 123} [\href{https://arxiv.org/abs/2301.07151}{{\ttfamily
  2301.07151}}].

\bibitem{Jenkins:2017jig}
E.~E. Jenkins, A.~V. Manohar and P.~Stoffer, \emph{{Low-Energy Effective Field
  Theory below the Electroweak Scale: Operators and Matching}},
  \href{https://doi.org/10.1007/JHEP03(2018)016}{\emph{JHEP} {\bfseries 03}
  (2018) 016} [\href{https://arxiv.org/abs/1709.04486}{{\ttfamily
  1709.04486}}].

\bibitem{Jenkins:2017dyc}
E.~E. Jenkins, A.~V. Manohar and P.~Stoffer, \emph{{Low-Energy Effective Field
  Theory below the Electroweak Scale: Anomalous Dimensions}},
  \href{https://doi.org/10.1007/JHEP01(2018)084}{\emph{JHEP} {\bfseries 01}
  (2018) 084} [\href{https://arxiv.org/abs/1711.05270}{{\ttfamily
  1711.05270}}].

\bibitem{Alvarez-Gaume:1981exv}
L.~Alvarez-Gaume and D.~Z. Freedman, \emph{{Geometrical Structure and
  Ultraviolet Finiteness in the Supersymmetric Sigma Model}},
  \href{https://doi.org/10.1007/BF01208280}{\emph{Commun. Math. Phys.}
  {\bfseries 80} (1981) 443}.

\bibitem{Nagai:2021gmo}
R.~Nagai, M.~Tanabashi, K.~Tsumura and Y.~Uchida, \emph{{Scalar and fermion
  on-shell amplitudes in generalized Higgs effective field theory}},
  \href{https://doi.org/10.1103/PhysRevD.104.015001}{\emph{Phys. Rev. D}
  {\bfseries 104} (2021) 015001}
  [\href{https://arxiv.org/abs/2102.08519}{{\ttfamily 2102.08519}}].

\bibitem{DeWitt:2012mdz}
B.~S. DeWitt, \emph{{Supermanifolds}}, Cambridge Monographs on Mathematical
  Physics. Cambridge Univ. Press, Cambridge, UK, 5, 2012,
  \href{https://doi.org/10.1017/CBO9780511564000}{10.1017/CBO9780511564000}.

\bibitem{rogers2007supermanifolds}
A.~Rogers, \emph{Supermanifolds: theory and applications}. World Scientific,
  2007.

\bibitem{Arnowitt:1975bd}
R.~L. Arnowitt and P.~Nath, \emph{{Riemannian Geometry in Spaces with Grassmann
  Coordinates}}, \href{https://doi.org/10.1007/BF00762016}{\emph{Gen. Rel.
  Grav.} {\bfseries 7} (1976) 89}.

\bibitem{tHooft:1973bhk}
G.~'t~Hooft, \emph{{An algorithm for the poles at dimension four in the
  dimensional regularization procedure}},
  \href{https://doi.org/10.1016/0550-3213(73)90263-0}{\emph{Nucl. Phys. B}
  {\bfseries 62} (1973) 444}.

\bibitem{Neufeld:1998js}
H.~Neufeld, J.~Gasser and G.~Ecker, \emph{{The one loop functional as a
  Berezinian}},
  \href{https://doi.org/10.1016/S0370-2693(98)00964-2}{\emph{Phys. Lett. B}
  {\bfseries 438} (1998) 106}
  [\href{https://arxiv.org/abs/hep-ph/9806436}{{\ttfamily hep-ph/9806436}}].

\bibitem{Henning:2016lyp}
B.~Henning, X.~Lu and H.~Murayama, \emph{{One-loop Matching and Running with
  Covariant Derivative Expansion}},
  \href{https://doi.org/10.1007/JHEP01(2018)123}{\emph{JHEP} {\bfseries 01}
  (2018) 123} [\href{https://arxiv.org/abs/1604.01019}{{\ttfamily
  1604.01019}}].

\bibitem{Buchalla:2017jlu}
G.~Buchalla, O.~Cata, A.~Celis, M.~Knecht and C.~Krause, \emph{{Complete
  One-Loop Renormalization of the Higgs-Electroweak Chiral Lagrangian}},
  \href{https://doi.org/10.1016/j.nuclphysb.2018.01.009}{\emph{Nucl. Phys. B}
  {\bfseries 928} (2018) 93}
  [\href{https://arxiv.org/abs/1710.06412}{{\ttfamily 1710.06412}}].

\bibitem{Buchalla:2019wsc}
G.~Buchalla, A.~Celis, C.~Krause and J.-N. Toelstede, \emph{{Master Formula for
  One-Loop Renormalization of Bosonic SMEFT Operators}},
  \href{https://arxiv.org/abs/1904.07840}{{\ttfamily 1904.07840}}.

\bibitem{Grzadkowski:2010es}
B.~Grzadkowski, M.~Iskrzynski, M.~Misiak and J.~Rosiek, \emph{{Dimension-Six
  Terms in the Standard Model Lagrangian}},
  \href{https://doi.org/10.1007/JHEP10(2010)085}{\emph{JHEP} {\bfseries 10}
  (2010) 085} [\href{https://arxiv.org/abs/1008.4884}{{\ttfamily 1008.4884}}].

\bibitem{Murphy:2020rsh}
C.~W. Murphy, \emph{{Dimension-8 operators in the Standard Model Eective Field
  Theory}}, \href{https://doi.org/10.1007/JHEP10(2020)174}{\emph{JHEP}
  {\bfseries 10} (2020) 174}
  [\href{https://arxiv.org/abs/2005.00059}{{\ttfamily 2005.00059}}].

\bibitem{Li:2020gnx}
H.-L. Li, Z.~Ren, J.~Shu, M.-L. Xiao, J.-H. Yu and Y.-H. Zheng, \emph{{Complete
  set of dimension-eight operators in the standard model effective field
  theory}}, \href{https://doi.org/10.1103/PhysRevD.104.015026}{\emph{Phys. Rev.
  D} {\bfseries 104} (2021) 015026}
  [\href{https://arxiv.org/abs/2005.00008}{{\ttfamily 2005.00008}}].

\bibitem{Fermi:1934sk}
E.~Fermi, \emph{{Trends to a Theory of beta Radiation. (In Italian)}},
  \href{https://doi.org/10.1007/BF02959820}{\emph{Nuovo Cim.} {\bfseries 11}
  (1934) 1}.

\bibitem{Dekens:2019ept}
W.~Dekens and P.~Stoffer, \emph{{Low-energy effective field theory below the
  electroweak scale: matching at one loop}},
  \href{https://doi.org/10.1007/JHEP10(2019)197}{\emph{JHEP} {\bfseries 10}
  (2019) 197} [\href{https://arxiv.org/abs/1908.05295}{{\ttfamily
  1908.05295}}].

\end{thebibliography}\endgroup

\end{document}